\documentclass{tcibook}
\usepackage{fancyhea}
\usepackage{work}
\usepackage{bm}       %    enables bold math symbols  e.g.  \bm{\gamma}
\usepackage{graphicx}
\usepackage{multirow}
\usepackage{lineno}
\usepackage{threeparttable}
\usepackage[normalem]{ulem}
\usepackage{color}
\usepackage[colorlinks = true,
            linkcolor = blue,
            urlcolor  = blue,
            citecolor = blue,
            anchorcolor = blue]{hyperref}

\usepackage[utf8]{inputenc}
\usepackage{authblk}

\def\beq{\begin{equation}}
\def\eeq#1{\label{#1}\end{equation}}
\def\eeqn{\end{equation}}

\newenvironment{Eqnarray}%
   {\arraycolsep 0.14em\begin{eqnarray}}{\end{eqnarray}}
\def\beqa{\begin{Eqnarray}}
\def\eeqa#1{\label{#1}\end{Eqnarray}}
\def\eeqan{\end{Eqnarray}}

\let\bar=\overbar

\def\lsim{\mathrel{\raise.3ex\hbox{$<$\kern-.75em\lower1ex\hbox{$\sim$}}}}
\def\gsim{\mathrel{\raise.3ex\hbox{$>$\kern-.75em\lower1ex\hbox{$\sim$}}}}

\def\del{\partial}
\def\Dslash{\not{\hbox{\kern-4pt $D$}}}
\def\dslash{\not{\hbox{\kern-2pt $\del$}}}
\def\pslash{\not{\hbox{\kern-2pt $p$}}}
\def\ETmiss{\not{\hbox{\kern-4pt $E$}}_T}

\def\Dlr{\mathrel{\raise1.5ex\hbox{$\leftrightarrow$\kern-1em\lower1.5ex\hbox{$D$}}}}

\def\MSB{{\bar{M \kern -2pt S}}}
\def\msb{{\bar{\scriptsize M \kern -1pt S}}}

\def\drb{{\bar{\scriptsize D \kern -1pt R}}}

\begin{document}

%%  uncomment this line to use line numbers in drafts:
%\linenumbers

\pagenumbering{roman}

\parindent=0pt
\parskip=8pt
\setlength{\evensidemargin}{0pt}
\setlength{\oddsidemargin}{0pt}
\setlength{\marginparsep}{0.0in}
\setlength{\marginparwidth}{0.0in}
\marginparpush=0pt

% The content begins here

\pagenumbering{arabic}

\renewcommand{\chapname}{chap:intro_}
\renewcommand{\chapterdir}{.}
\renewcommand{\arraystretch}{1.25}
\addtolength{\arraycolsep}{-3pt}

%%%%%%%%% Styling of CEF Topical Group report "Recommendations"
%%%%%%%%%%%%%%%%%%%%%%%%%%%%%%%%%%%%%%%%%%%%%%%%%%%%%%%%%%%%%%%%

%%%%%% USAGE %%%%%%
%%%%%%%%%%%%%%%%%%%

%% Once for each topical report, edit the below command to make it specific to your topical group
%% i.e. change the 6 to whatever group numbers yours is!
\newcommand{\cefgroup}{1}

%% Include the below syntax somewhere in your main .tex file
%% \input{CEF-recommendation-style}
%% (or else otherwise include the commands and environment defined here somewhere)

%% For each recommendation, use the following syntax:
%
% \begin{recs}
%     \rec{A}{B}{C}
% \end{recs}
% 
% A: The number of the recommendation (e.g. 6)
% B: The concise text of the recommendation (e.g. Breese should take a vacation.)
% C: Any supporting text (e.g. Breese has been working very hard for a very long time and has earned a break!)
%
%% If you want to make multiple recommendations within the same pair of thick horizontal bars,
%% Use the \rec command multiple time separated by blank lines
%
% \begin{recs}
%     \rec{A}{B}{C}
%
%     \rec{D}{E}{F}
%
%     \rec{G}{H}{I}
% \end{recs}

\newenvironment{recs}
    {\noindent \begin{minipage}{\textwidth} \rule{\textwidth}{2mm}}
    {\rule{\textwidth}{2mm} \end{minipage}}

\newcommand{\rec}[3]{
\nobreak \noindent \textbf{CEF0\cefgroup~Recommendation #1 -- #2}\\ 
\nobreak \rule{\textwidth}{0.4mm}
%\textit{#3}\\
\nobreak #3 \\

}

\setcounter{chapter}{10} 

\chapter{Community Engagement Frontier}

\vskip0.2in

\begin{center} {\bf Frontier Conveners:} {Kétévi~A.~Assamagan$^1$, Breese~Quinn$^2$}\end{center}

\begin{center} {\bf Topical Group Conveners:} {Kenneth~Bloom$^3$, Véronique~Boisvert$^4$, Carla~Bonifazi$^5$, Johan~S.~Bonilla$^6$, Mu-Chun~Chen$^7$, Sarah~M.~Demers$^8$,  Farah~Fahim$^9$, Rob~Fine$^{10}$, Mike~Headley$^{11}$,  Julie~Hogan$^{12}$, Kathryn~Jepsen$^{13}$,  Sijbrand~de~Jong$^{14}$, Aneliya~Karadzhinova-Ferrer$^{15}$, Yi-Hsuan~Lin$^{16}$, Don~Lincoln$^9$,  Sudhir~Malik$^{17}$, Alex~Murokh$^{18}$,  Azwinndini~Muronga$^{19}$, Randal~Ruchti$^{20}$,  Louise~Suter$^9$, Koji~Yoshimura$^{21}$}\end{center}

\begin{center} {\bf Other Contributors:} {Erin~V.~Hansen$^{22}$, Samuel Meehan$^{23}$, Erica~Smith$^{24}$} \end{center}

\begin{center}
$^1${Department of Physics, Brookhaven National Laboratory, Upton, NY, 11973, USA} \\
$^2${Department of Physics and Astronomy, University of Mississippi, Oxford, MS, 38677, USA}  \\
$^3${Department of Physics and Astronomy, University of Nebraska-Lincoln, Lincoln, NE 68588, USA} \\
$^4${Department of Physics, Royal Holloway University of London, Egham Hill, Egham Surrey, TW20 0EX, United Kingdom} \\
$^5${ICAS-ICIFI-UNSAM/CONICET, Argentina, and Universidade Federal do Rio de Janeiro, Brazil} \\
$^6${Department of Physics and Astronomy, University of California, Davis, CA, 95616, USA} \\
$^7${Department of Physics and Astronomy, University of California, Irvine, CA, 92697, USA} \\
$^8${Department of Physics, Yale University, New Haven, CT, 06511, USA} \\
$^9${Fermi National Accelerator Laboratory, Batavia, IL, 60510, USA} \\
$^{10}${Los Alamos National Laboratory, Los Alamos, NM, 87545, USA} \\
$^{11}${Sanford Underground Research Facility, Lead, SD, 57754, USA} \\
$^{12}${Department of Physics \& Engineering, Bethel University, St Paul, MN, 55112, USA} \\
$^{13}${Symmetry Magazine, SLAC National Accelerator Laboratory, Menlo Park, CA, 94025, USA} \\
$^{14}${Faculty of Science, Radboud Universiteit, 6525 AJ Nijmegen, Netherlands} \\
$^{15}${Helsinki Institute of Physics (HIP) P.O. Box 64, 00014 University of Helsinki, Finland, and Lappeenranta University of Technology (LUT), School of Engineering Science, Box 20, 53851 Lappeenranta, Finland} \\
$^{16}${Queen’s University, Department of Physics, Engineering Physics \& Astronomy, Kingston ON, Canada}  \\
$^{16}${SNOLAB, Creighton Mine \#9, 1039 Regional Road 24, Sudbury ON, Canada}  \\
$^{17}${Physics Department, University of Puerto Rico-Mayaguez, Mayaguez, PR, 00681, USA} \\
$^{18}${RadiaBeam Technologies, Santa Monica, CA, 90404, USA} \\
$^{19}${Faculty of Science, Nelson Mandela University, Gqeberha, South Africa} \\
$^{20}${Department of Physics and Astronomy, University of Notre Dame, Notre Dame, IN, 46556, USA} \\
$^{21}${Department of Physics, Okayama University, Okayama, 700-8530, Japan} \\
$^{22}${Department of Physics, University of California, Berkeley, Berkeley, CA, 94720, USA} \\ 
$^{23}${AAAS Science \& Technology Policy Fellow, Washington, DC, 20005, USA} \\
$^{24}${Department of Physics, Indiana  University, Bloomington, IN, 47405, USA}
\end{center}

\tableofcontents

\newpage

\section{Introduction}
\label{sec:intro}

In Snowmass 2013, investigations into HEP community engagement addressed physics Communication, Education and Outreach (CE\&O)~\cite{https://doi.org/10.48550/arxiv.1401.6119}, and this was the first time HEP incorporated such activities into the Snowmass process. The CE\&O Frontier organized its Working Groups according to the target audiences of the General Public, Policy Makers, Science Community, and grade 5-12 Teachers and Students. The framework for all CE\&O efforts consisted of three main outcome goals: ensuring the resources needed for the US to maintain a leadership role in HEP research, ensuring the public realizes how valuable and exciting particle physics is, and ensuring US HEP produces a talented and diverse pool of STEM professionals. Common themes of action to meet those goals emerged across the Working Groups: making a coherent and unified case for HEP, instituting real recognition for colleagues engaged in CE\&O activities, providing more CE\&O resources and training for our community, and creating a central team tasked with supporting HEP CE\&O efforts.

Each Snowmass 2013 CE\&O Working Group produced a number of specific recommendations for implementation to help achieve the CE\&O goals. In the intervening years since the release of the CE\&O Frontier Report, many individuals and institutions accomplished much in following several of those recommendations. Great examples can be found in the area of public policy. A small group of people developed sophisticated, powerful and efficient database and wiki tools that have transformed and multiplied our HEP Congressional advocacy efforts. Another major communications success for the HEP community during this last P5 era has been the development and consistent delivery of a coherent, unified message about the US HEP program to Congress and other audiences. On the other hand, our field essentially set aside many of those recommendations. The community did not create a central or national team to support CE\&O work, nor did it put in place many institutional incentives to encourage that work among our colleagues. Although much good work was done by many, implementation of CE\&O recommendations has been spotty and incoherent overall. 

The Snowmass 2013 experience absolutely did make clear that it is crucial to address community engagement for the health of US HEP. In fact, leadership within the field realized from the very beginning of the Snowmass 2021 process that community engagement needed to be expanded into a Community Engagement Frontier (CEF) with full scope over all areas of community engagement (i.e. a much broader scope than CE\&O from 2013), and co-equal with the physics frontiers. The structure of this expanded Community Engagement Frontier includes seven Topical Groups defined primarily by general issues to be addressed: Applications and Industry; Career Pipeline and Development; Diversity, Equity and Inclusion; Physics Education; Public Education and Outreach; Public Policy and Government Engagement; and Environmental and Societal Impacts. The overall objective is to improve and sustain strategic engagements with our communities in order to draw support for and strengthen the field of particle physics (an inward-focused goal carried forward from Snowmass 2013), while playing key roles in serving those communities (an outward-focused goal added for Snowmass 2021). Arranging the Topical Groups by issue proved to be a very efficient structure that brought great focus to the efforts of the Groups which, succinctly, aim to support: practical applications of research in particle physics and technology transfers to industries; career development and job opportunities for young scientists; encouragement and inclusion of diverse physicists to reflect the diversity in our communities; advances in physics education to produce talented and qualified students; engagement with the public to share in the essence and importance of physics research; partnerships with governments and policymakers to grow the scientific enterprise; and improvements in the ways that our field affects the environment and society in which we live. 

The issues addressed by the CEF Topical Groups are relevant to all the other frontiers of Snowmass 2021 — they are crosscutting topics or issues for all the physics frontiers. About one hundred letters of interest (LOI) were received on these topics. These LOIs were condensed into thirty-five contributed papers, the details of which are mentioned in the Topical Group Reports. In addition to the LOIs, inputs to contributed papers came from town hall discussions, group meetings, expert speaker invitations, workshops, and surveys. Sections ~\ref{sec:CEF01}--\ref{sec:CEF07} of this report summarize the work done by the Topical Groups, and the individual Topical Group Reports for CEF01~\cite{https://doi.org/10.48550/arxiv.2210.01248}, CEF02~\cite{https://doi.org/10.48550/arxiv.2209.10114},  CEF03~\cite{https://doi.org/10.48550/arxiv.2209.12377}, CEF04~\cite{https://doi.org/10.48550/arxiv.2209.08225}, CEF05~\cite{https://doi.org/10.48550/arxiv.2210.00983}, CEF06~\cite{https://doi.org/10.48550/arxiv.2209.09067} and CEF07~\cite{https://doi.org/10.48550/arxiv.2209.07684} document the details and the specific recommendations for improving the ways our community engages in these areas.

\subsection{Overall Community Engagement Frontier Goals}

\begin{figure}[!htbp]
\begin{center}
\includegraphics[width=0.5\hsize]{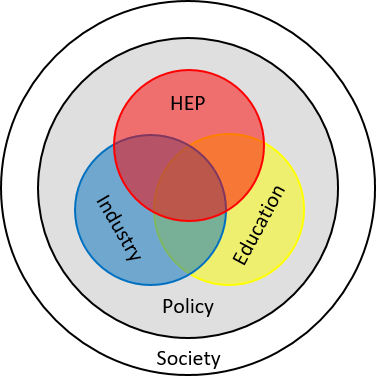}
\hfill
\end{center}
\caption{Five interrelated communities targeted for HEP engagement.}
\label{fig:Communities}
\end{figure}

Structuring the CEF work into the seven issue-based Topical Groups defined above was a very effective strategy for organizing and maximizing the productivity of the people working in the Frontier. However, two categories of overlaps necessitate a different organizing principle for setting overall implementation goals for CEF. First, the shared experiences and overlapping interests among the various Topical Groups and other frontiers mean that different groups are often addressing the same issue from different directions or perspectives, and we need to continue following the guiding principle of coherence to our efforts. Second, there are often two or more very different recommendations related by a common target audience. Bringing those recommendations together could very well result in more efficient and effective engagement.  These considerations informed the development and incorporation of strategies and recommendations articulated in a set of overarching goals for HEP engagement with five interrelated communities: HEP itself, K-postdoc education, private industry, government policy, and the broader society (Figure~\ref{fig:Communities}). These overall CEF goals organized by target community for engagement actually echo the Snowmass 2013 working group structure. The goals along with references to the specific Topical Group sections informing each goal are listed below.
%%%%%%%%%%%%%%%%%%%%%%%%%%%%%%%%%%%%%%%%%%%%%%%%%%%%%%%%%%%%%%%%%%%%%%%%%
%%%%%%%%%%%%%%%%%%%%%%%%%%%%%%%%%%%%%%%%%%%%%%%%%%%%%%%%%%%%%%%%%%%%%%%%%%%

\subsubsection*{HEP Internal Engagement}

It is often said that you can tell much about a group of people by considering how they interact with each other and conduct their own affairs. Therefore, a reasonable place to begin a study of HEP Community Engagement is to analyze how the HEP community engages with itself. What are the characteristics of the ways that we build and organize our own community of colleagues? What values do we embody by our choices of which activities are incentivized (or not) in our work? Every single CEF Topical Group confronted these questions at varying levels. In fact, a plurality of recommendations put forward by the seven Groups were directed inward, suggesting improvements to the standard operating procedures of the US particle physics community.

It is widely known that science broadly is a discipline which lags behind most others in its membership  diversity along multiple axes. Physics, and HEP in particular, have been less successful than most other science fields in realizing much improvement in this area over the years. It has become clear that a factor contributing to this limited success is the fact that the pool from which we draw professional talent is a small one dominated by traditional physics programs at a select group of R1 academic institutions. This must change for HEP to access the depth of talent from the broader society. %needed for our projects and programs to succeed. 
At the same time, individuals who do bring greater diversity to our field often encounter barriers to full participation and advancement in their careers. The norms of interaction developed over decades by a fairly homogeneous community can serve to alienate those possessing a potential to enrich our field with different backgrounds and perspectives. It is not only beneficial, but also simply good manners to present a welcoming environment to new colleagues and neighbors.

Most particle physicists are receptive to participation in Community Engagement activity, and many are quite active in this work. However, there are strong pressures within HEP that serve to prevent many members, especially early career members, from significant participation. The first of these is time. HEP research is a demanding task requiring great resources, not the least of which is time. There are always schedules to follow, deadlines to meet, tasks to complete. Particularly for postdocs and junior faculty working to establish themselves in the field, it is a tall order to expect them to sacrifice research time for community engagement efforts. The competition for the next job or promotion is fierce, and time “lost” equates to falling behind one’s peers in career achievement. This leads to the second pressure, which is the fact that records of successful participation in community engagement tasks have rarely been given strong consideration by hiring or promotion evaluation committees at our labs and universities. There are signs that this may be changing, as cases of work such as “outreach” being included in the “service” portion of committee evaluations is more common than it once was, and NSF has long encouraged attention to similar efforts through its grant requirement to address broader impacts.

In the P5 era since Snowmass 2013, US HEP has become renowned for its record of project management. Until now, that success has primarily relied on maintaining a proper balance of our projects’ scientific capabilities, budgets and schedules. Over the past decade, the worldwide HEP community has come to realize that another concern must be added to this balancing act: the direct impacts that our activities have on the communities and environment in which we exist. This means that we must plan to limit the specific and sometimes unique impacts that our collaborative projects and individual work have on the climate and broader environment.
These and other aspects of internal HEP community engagement resulted in the formation of many specific recommendations for changes or improvements within our field, all of which can be found in the various CEF Topical Group Reports. Overlaps and relationships that exist among the inward-directed recommendation of each Topical Group led to the development of the following goals for engagement within the HEP community (each referenced by the Topical Groups whose reports most directly relate to that goal).

\begin{itemize}
    \item {\bf The HEP community should institute a broad array of practices and programs to reach and retain the diverse talent pool needed for success in achieving our scientific vision. In particular, we need to encourage stronger participation in HEP collaborations by faculty and students from non-R1 academic institutions. (CEF02: Section~\ref{sec:CEF02}; CEF03: Section~\ref{sec:CEF03}; CEF04: Section~\ref{sec:CEF04})} 
    \item {\bf HEP communities are still plagued by the alienation experienced by marginalized physicists who are part of the community. HEP needs to address these persistent issues by employing the use of robust strategic planning procedures including a full re-envisioning of our workplace norms and culture to prioritize eliminating the barriers and negative experiences faced by our marginalized colleagues. (CEF03: Section~\ref{sec:CEF03})} 
    \begin{itemize}
        \item {\bf Research institutes and universities should do more to maintain the highest standard in work-life balance and mental health of staff. Proper training of staff should be developed to integrate productive work habits that encourage a balance between professional expectations and private affairs, and good mental health. (CEF02: Section~\ref{sec:CEF02}; CEF03: Section~\ref{sec:CEF03}; CEF04: Section~\ref{sec:CEF04})}
    \end{itemize}
    \item {\bf The HEP community needs to address under-representation of many groups within the field by implementing new modes of community organization and decision-making procedures that promote agency and leadership from all stakeholders within the scientific community. (CEF03: Section~\ref{sec:CEF03})} 
    \begin{itemize}
        \item {\bf Funding agencies, Universities, laboratories, and HEP groups should improve and sustain international outreach, partnerships, schools, workshops, conferences, training, short-visits for research, and development of research consortia; mechanisms should be developed to facilitate the participation of colleagues from developing countries.  (CEF03: Section~\ref{sec:CEF03})}
    \end{itemize}
    \item {\bf In addressing the unique needs and issues of marginalized physicists, HEP communities must engage in partnership with scholars, professionals, and other experts in several disciplines, including but not limited to anti-racism, critical race theory, and social science. (CEF03: Section~\ref{sec:CEF03})} 
    \item {\bf All HEP communities should create structures to fully open career path opportunities to everyone, and to conduct event planning to ensure events are accessible to all community members, especially those with disabilities. (CEF03: Section~\ref{sec:CEF03})} 
    \item {\bf The HEP community should enact structural changes to foster broader, deeper, and more effective participation in community engagement, through policies such as considering community engagement work in hiring, promotion, and grant decisions. (CEF05: Section~\ref{sec:CEF05}, et.al.)}
    \begin{itemize}
        \item {\bf Individual scientists should encourage others, including peers, mentees, and students, by participating in public engagement and discussing its importance. (CEF05: Section~\ref{sec:CEF05})}
    \end{itemize}
    \item {\bf HEP needs to improve communication channels with the funding agencies and internal communication within HEP concerning program planning and funding, particularly with regard to the upcoming P5 plan. (CEF06: Section~\ref{sec:CEF06})}
    \item {\bf HEP must take greater responsibility for its impacts on climate change by addressing and mitigating these impacts through DOE project policies and individual community member actions. (CEF07: Section~\ref{sec:CEF07})}
\end{itemize}

\subsubsection*{Education}

Professionals in particle physics must be prepared for careers in the field through instruction in the skills, techniques and investigation processes characteristic of the discipline. This training primarily occurs within the education community spanning kindergarten through postdoc (K-PD). The development of foundational skills and interests must begin in early grades of local schools, and should eventually expand to include learning experiences in international settings.  To produce colleagues with the specific abilities required for modern HEP research, our academic institutions need to be teaching content that matches the needs of our discipline. However, US universities for example, are sometimes towers of tradition, slow to adapt to changing career environments, especially when people active in the fields do not communicate the needs for change. %However, US universities for example, are sometimes towers of tradition, slow to adapt to changing career environments, especially when those of us active in those careers are not engaging with the institutions to communicate those current needs. 

One particular area of disconnect is the trend that has developed of US HEP becoming a more specialized pursuit, weighted more and more heavily toward the academic, analytical side of the work. With changing dynamics in the organization and funding of HEP within the US, many HEP individuals and collaborations have less familiarity with the technical or engineering expertise required for our projects. In addition, modern particle physics analysis depends on the application of specific skills that are not often part of the standard degree programs in many universities. These can range from tried-and-true statistics that are often learned “on-the-job” in our field, to more novel disciplines such as Artificial Intelligence algorithms that are becoming increasingly common in our work. 

These concerns along with needs to improve connections with K-12 and international students are reflected in goals for HEP engagement with the Education community, formulated to address educational deficiencies specific to our field. 

\begin{itemize}
    \item {\bf Our field cannot absorb all the early career members that it produces, so funding agencies, national laboratories and universities should work together to provide more education and career opportunities for engineering and industry-focused research within and outside HEP, and update degree programs to match better the skills needed and career opportunities required in today's HEP and related fields. (CEF01: Section~\ref{sec:CEF01}; CEF02: Section~\ref{sec:CEF02}; CEF04: Section~\ref{sec:CEF04})} 
    \item {\bf HEP academia should work with K-12 teachers and students to create supportive local communities to nurture student interest in math and science. (CEF04: Section~\ref{sec:CEF04})} 
    \item {\bf Pre-university and university programs for international student collaboration need to be expanded and supported, especially to partner with colleagues in developing countries. (CEF04: Section~\ref{sec:CEF04})} 
\end{itemize}

\subsubsection*{Industry}

In times past, the US HEP community was supported by strong direct relationships with vibrant industries well matched to the technological needs of the field. Over the years, as modes of project management, funding, licensing etc. have evolved, it has become increasingly difficult for US HEP to provide support through project partnerships adequate to support a viable industrial base of US companies capable of providing the technology production required for successful execution of HEP projects. Often, these mutually beneficial two-way relationships remain stronger in Europe and Asia than they do in the US. For example, industrial accelerator companies in the US represent a rather small community. 

Improvements in HEP-industry relationships can be achieved at all scales. HEP could strengthen relationships with large microelectronic firms through new models of agency- or field-wide licenses and platform access. Partnerships with smaller startup companies often exist and are much easier to nurture at the lab or university level, so those connections should be leveraged. From the industry perspective, the reduction of cross-agency or cross-office barriers will enhance and accelerate innovation.

Many of the proposals envisioned for promoting strong relationships between the Industry community and HEP are represented in the following overarching goals for Industry engagement.

\begin{itemize}
    \item {\bf Funding agencies and the national laboratories should enhance policies and programs to promote a more fertile environment for cross-agency technology development and technology transfer, and to support co-development of specific technologies with industries such as accelerators, microelectronics, and FLASH-RT. (CEF01: Section~\ref{sec:CEF01}; CEF02: Section~\ref{sec:CEF02})}
    \item {\bf Together, laboratories and universities need to help bolster our US industrial support base by pursuing targeted partnerships with early stage scaleup companies on HEP projects. (CEF01: Section~\ref{sec:CEF01})}
    \item {\bf The HEP community needs to strengthen ties with industry and other fields by developing effective alumni networking tools and programs, to facilitate transitions to industry careers and encourage industry collaboration on HEP projects. (CEF02: Section~\ref{sec:CEF02})}
\end{itemize}

\subsubsection*{Policy}

When we speak of Policy in this context, we do not refer to the official rules themselves by which an organization operates. We refer to a specific target community for HEP engagement. That is the Policy community, or collection of various governments and individuals working within government that have influence on enacted policies that directly affect our field. HEP, Education, and Industry communities all operate and exist within a milieu of government policy.

The core of HEP’s Policy engagement has always been its highly effective Congressional advocacy, conducted by the major Users groups representing US particle physics to secure strong funding for US HEP. The primary component of that Congressional advocacy is the “DC Trip,” which has been developed, expanded and refined by a small group of dedicated colleagues for decades. It has become a major sophisticated operation that is held up to other fields as the gold standard of scientific advocacy. Throughout the P5 era, HEP has successfully garnered strong bipartisan support in Congress for the field’s program of research. One weakness of this effort is the wholly volunteer nature of the stewardship of our Congressional advocacy. The HEP community should dedicate resources to put it on a sustainable path for future growth.

On the other hand, there is considerable room for improvement in our advocacy to the federal Executive branch. Across most administrations, support for HEP in Presidential budget proposals has typically not reached the level of Congressional support for many years. Our interactions with the Executive branch are limited in breadth and frequency, so this is a potential area of significant growth. The HEP community has rarely, if ever, mounted any real effort to advocate on behalf of the field to state or local government. While state and local advocacy would likely be selective in its application, it still represents a real growth opportunity for HEP support.

Colleagues have expressed interest in advocating for non-HEP funding issues that may not be HEP-specific, but have direct impact on our field and its health. Examples include issues such as VISA and immigration policy. The Users group Congressional advocacy has never directly promoted policies outside of HEP funding, and likely would not attempt to advocate for non-HEP-specific issues on its own. The main reason being the general principle that advocacy is best carried out by the largest group available with common purpose on a given issue. Therefore, partnership with broader scientific societies such as APS, AAAS, etc., would be a more powerful form of HEP advocacy for more general scientific issues.

Consequently, the greatest growth in HEP Policy engagement would be achieved by achieving the following overall goals.

%\begin{itemize}
%5   \item  APS DPF, HEPAP, and the user groups need to review the structure of who, in the community, is responsible for advocacy;
%    \item  An option of forming an HEP community government engagement group should be considered, with the responsibility of expanding government engagement capabilities.
%\end{itemize}

\begin{itemize}
    \item  {\bf APS DPF, HEPAP, and the user groups need to review the structure of HEP advocacy, including considering the formation of an HEP community government engagement group with  responsibility for expanding government engagement capabilities. (CEF06: Section~\ref{sec:CEF06})} \item {\bf The HEP user groups and DPF need to provide the resources for continued growth and sustainability of the annual HEP Congressional advocacy effort. (CEF06: Section~\ref{sec:CEF06})}
    \item {\bf The users groups, DPF, laboratories and universities should build on our successful HEP Congressional advocacy by expanding our advocacy to the federal executive branch and state and local governments. (CEF06: Section~\ref{sec:CEF06})}
    \item {\bf HEP should establish a group in partnership with other science and physics societies on advocacy for non-HEP funding issues. (CEF06: Section~\ref{sec:CEF06})}
\end{itemize}

\subsubsection*{Society}

Each of the four target communities described above are subsets of the broader society at large, existing within what is sometimes referred to as the general public. When we speak of broader society as a target community for engagement, typically we are referring to the portion of society that is the complement of the union of the other four communities. In other words, everyone in society outside of HEP, Education, Industry, and Policy.

The key theme in Societal engagement is reflected in the term engagement itself. Until recently, interactions with the public were usually spoken of as Public Outreach. However, outreach implies reaching out to some group. In other words, it is a one-way activity between groups that are not on the same level. HEP is telling other people something. What we’re saying might be good, but we’re not listening. This is not an effective means of building relationships. Conversely, most groups involved in outreach communication have evolved to frame what they do as engagement, or engaging with another group. This implies a co-equal partnership, hopefully one that is mutually beneficial.

This value leads to the articulation of two major goals for real community engagement with the broader society, one very general, and one very specific.

\begin{itemize}
    \item {\bf HEP needs to transition from an ethos of conducting outreach and communication to the public, to a culture of engagement in relationships with the public. This should be done by building lasting relationships with the full breadth of all of our supporting communities (especially those that have been historically excluded) that are not based on transactional interactions, but rather real two-way partnerships that consider the needs and interests of the audience and include its members in program design. (CEF05: Sections~\ref{sec:CEF05}; CEF07: Section~\ref{sec:CEF07})}
    \item {\bf HEP should build synergistic collaborations with the non-proliferation community that draw on a broader spectrum of funding sources for work on HEP-specific technologies related to nuclear non-proliferation. (CEF07: Section~\ref{sec:CEF07})} 
\end{itemize}

\subsection{Participation and Implementation}

Throughout the Snowmass~2021 process, there has been relatively low participation in Community Engagement Frontier efforts. This is true although it is generally agreed that CEF topics are cross cutting, i.e. they affect the entire community to various degrees, and thus are important to be addressed. The vast majority of CEF work was carried out by the small number of frontier and topical group conveners, with only a few additional dedicated community members making significant contributions. Indeed, many contributed paper study groups were led and conducted by the CEF topical group conveners themselves through to publications of these papers, because of low community involvement. Instead of CEF being a cross-cutting frontier with significant participation, it became an isolated set of activities---in spite of frontier liaisons---carried out by a relatively small group of people; almost all of whom are also physicists with interest in the other “physics” frontiers. These dedicated volunteers were largely prevented from participating in the Snowmass physics frontiers at significant levels due to the burden placed on them by the lack of participation in CEF by the majority of their HEP colleagues. As a result, all of the CEF topical group conveners sacrificed career advancement opportunities in order to carry out the work necessary for the health of the field on behalf of the entire HEP community. Various reasons have been suggested to explain low involvement in CEF, e.g. lack of time or the fact that career progression depends on the quality of research output rather than community engagement effort. Surveys done in Refs.~\cite{https://doi.org/10.48550/arxiv.2203.07328, https://doi.org/10.48550/arxiv.2210.00983} offer further insights into the low participation issue. It is also possible that Snowmass is the wrong time for the community to focus on developing a plan for addressing CEF issues. Certainly, individuals’ concerns for time management and maximizing the potential for career development are heightened during the high-stakes planning and decision-making of Snowmass. Snowmass may be the one time that presents the greatest barriers to a large fraction of the community choosing to participate in CEF, so perhaps this should be the last time that CEF is a part of Snowmass. \emph{Whatever the reasons, the HEP community both corporately through structural change and as individuals through personal reflection must decide that everyone’s participation in CEF issues is required for the field to become healthy and grow. We can no longer rely on a small number of colleagues to shoulder the full burden of this work to the detriment of their own careers. Our field simply will not survive otherwise.}

If the field of HEP does decide that CEF issues are worth addressing, and furthermore commit to doing so, then a plan of implementation of CEF recommendations must be developed. In the past, the responsibility for guiding the selection and implementation of Snowmass recommendations has been wholly delegated to the Particle Physics Project Prioritization Panel (P5). This arrangement has worked exceedingly well over the last 8 years for implementing a consensus plan of HEP projects. However, experience has shown that the current P5 mandate and makeup is not suited to adequately shepherd other areas of the HEP enterprise, including community engagement issues. For example, Snowmass 2013 included the Community, Education, and Outreach Frontier, which produced a report that made several recommendations~\cite{https://doi.org/10.48550/arxiv.1401.6119}. Neither P5 nor any other HEP organization or leaders took ownership of those issues to ensure that recommendations from that frontier were implemented. As a result, despite the fact that individual people and institutions in our community have made efforts in these areas in the last decade, little overall progress on these recommendations has been realized since the last Snowmass.

To avoid a similar fate for Snowmass~2021 CEF recommendations, US HEP must establish a structure by which designated entities are given ownership of and responsibility for ensuring that CEF recommendations are implemented and monitored for progress. One such possibility is simply expanding the P5 charge to encompass CEF issues. It is not clear that is the most appropriate solution, though. P5 was explicitly designed to effectively prioritize (largely experimental) HEP projects. It may be that P5 is not ideally situated in the HEP ecosystem, or appropriately staffed to evaluate, choose, and monitor the progress of CEF initiatives. Considering the five engaged communities around which the CEF goals are organized, it could be that the American Physical Society Division of Particle and Fields (DPF) is best positioned to shepherd the internally-focused recommendations for HEP because that organization is most representative of the entire field’s membership. Perhaps a group formed in partnership between the funding agencies, the laboratories, and universities is needed to manage engagements with industry since that is where most of the direct relationships with industry are formed. Universities Research Association’s membership consisting of university administrations and its role in connecting academia with laboratories could make it best suited as a sponsoring organization for a team to work towards implementation of education initiatives. An option of forming an HEP community government engagement group, composed of elected community representatives and policy experts, to expand advocacy efforts, should be considered. %Almost certainly, the elected representative user groups that have led HEP funding advocacy efforts for decades should form the core of any new entity formed to expand policy efforts. 
As the focal points for public engagement with HEP, the laboratories themselves may be the ideal choice to manage the recommended programs directed at the broader society. \emph{All of these stakeholders should begin the conversation that must lead to an agreed upon structure for taking responsibility for implementing CEF recommendations. If this does not happen, then we face another decade of no progress on CEF issues in the HEP community.}

The rest of this report describes the impressive amount of high-quality work that a small number of your colleagues accomplished on your behalf. We hope that as you read and consider this content you will be convinced of two propositions:
\begin{enumerate}
    \item {\bf It is critical that we as individuals agree on the importance of all working together to address CEF issues in HEP.}
    \item {\bf A structure within HEP for taking ownership and responsibility for implementing CEF recommendations and monitoring their progress must be developed.}
\end{enumerate}
If we as a field can make these two ideas a reality, then US HEP will be much stronger and healthier by the time we gather for the next Snowmass process.

\section{CEF01: Applications and Industry}
\label{sec:CEF01}

\renewcommand{\cefgroup}{1}

The charge for the topical group CEF01: Applications and Industry is to develop strategies to strengthen HEP/Industry relationships in both directions, i.e. forming more partnerships to draw on industry expertise to further HEP goals and building on programs to facilitate transfer of HEP technologies/techniques for use in the broader society. This group considered the relationship between HEP laboratories, universities, and industrial stakeholders. In particular, CEF01 pursued the following objectives: (1) how to create an innovation ecosystem mutually beneficial to national laboratories, academia, and industry, (2) how to maximize the HEP-funded technology outcomes benefit to practical applications, (3) how to encourage co-development of related applications across agencies and programs, and (4) how to leverage HEP project partnerships to enable innovators to become entrepreneurs through tech commercialization.

In order to expand the discovery reach of experimental high energy physics, innovations in a variety of technologies are required to push operational and measurement tools and techniques to ever-higher levels of spatial and temporal precision. Not only do these advances propel scientific discovery, but they also enhance industrial capabilities to deliver novel and powerful applications to the benefit of the broader society. Strong development relationships between laboratories and universities of the HEP community and small to large scale tech companies in the industrial community are key for building an efficient and sustainable ecosystem for advancing technologies such as accelerators, microelectronics, artificial intelligence, and quantum information.

With individual modern HEP experiments characterized by industrial scales such as detectors with more than 1 billion sensors, 70 kilotons of liquid argon, or data rates equivalent to the entire North American internet traffic, it is obvious that a robust and diverse array of industry partners is necessary for HEP to mount almost any project in its portfolio. On the other hand, the engineering design benchmarks needed to handle the extreme radiation, cryogenic, low power, and inaccessible operating environments of HEP projects often exceed those typically found in industry by orders of magnitude. Multidisciplinary technology design and production partnerships for HEP between national labs, academia, and industry accelerates lab to fab innovation, prototyping to scale, technology maturation, spin-off development, and rapid adoption, all of which benefits industry capabilities; it also accelerates scientific discovery across the landscape of federally-funded research.

Barriers to effective HEP-Industry partnerships do exist. Although Small Business Innovation and Research (SBIR) funding can facilitate lab and university partnerships with small business, SBIR timeframes and funding levels are inadequate to support the large-scale HEP projects that require collaboration with big business. HEP technology goals and requirements are often not communicated to industry broadly and effectively. Economy of scale with regard to tool and license purchases from industry is typically not exploited in HEP. Science goals are not well-mapped to technology goals across funding agency offices, often resulting in the loss of synergistic co-development opportunities.

Those working in CEF01 produced five contributed papers examining three different modes of collaborative partnership, and cooperation on three different specific areas of technology:
\begin{itemize}
    \item Programs enabling deep technology transfer from national labs~\cite{https://doi.org/10.48550/arxiv.2203.15128};
    %\item Technology transfer with scaleups;
    \item Application-driven engagement with universities, synergies with other funding agencies~\cite{https://doi.org/10.48550/arxiv.2203.14706};
    \item Big industry engagement to benefit HEP: microelectronics support from large CAD companies~\cite{https://doi.org/10.48550/arxiv.2203.08973};
    \item Transformative technology for FLASH radiation therapy~\cite{https://doi.org/10.48550/arxiv.2203.11047};
    \item Nurturing the industrial accelerator technology base in the US~\cite{https://doi.org/10.48550/arxiv.2203.10377};
\end{itemize}
The following sections summarize the ideas and set forth the suggestions arising from each of those papers, and detailed in Ref.~\cite{https://doi.org/10.48550/arxiv.2210.01248}.

\subsection{Programs Enabling Deep Tech Transfer from National Labs}
To achieve the scientific goals of HEP projects, DOE national laboratories and the experiments they host require innovative ideas and at-scale prototyping for novel technologies. These projects must bridge the academy’s drive to push beyond state-of-the-art capabilities to industry’s forte in quality control and reliability. Experiments and industry collaborating to move these novel technologies from ideas to robust and cost-effective mid-scale manufacturing, can put industry on a path to commercial production. However, a proper environment is necessary to nurture this laboratory to fabrication (or lab to fab) technology transfer process in a manner that can sustain spin-off development and startup ventures over the long term. Several recommendations are made to develop an effective technology transfer ecosystem or for HEP at the national labs:

\begin{itemize}
    \item {\bf DOE should implement specific policy changes to foster deep technology transfer. (CEF01 Recommendation 4)}
\end{itemize}

\begin{quote}
    There are several specific policies within DOE that could be optimized to encourage more efficient technology transfer. Aligning inventor royalty distribution consistently across the DOE complex would simplify commercialization of technology developments. Laboratory and/or Division royalty shares might be used to support additional projects, rewards programs, and innovation/entrepreneurship educational opportunities for laboratory staff.
    
    The use of Partnership Intermediaries (PI) can accelerate commercialization, particularly for laboratories supporting HEP facilities with limited Technology Transfer (TT) resources. A PI is a non-profit with specialized skills to assist federal agencies and laboratories in TT and commercialization. Past pilot programs in the Department of Energy Office of Technology Transitions have shown promise in assessing technologies for market pull, marketing HEP-related technologies, and matchmaking technologies at national labs with entrepreneurs in private industry. This can be particularly successful by identifying “dual-use” applications early, which can be leveraged to enable upfront marketplace analysis and speed market acceptance.
    
    Other Transaction Authority (OTA) is a special mechanism federal agencies use to obtain or advance R\&D or prototypes. The government’s procurement regulations and certain procurement statutes do not apply to OTs; thus, OTA gives agencies flexibility to develop agreements tailored to a particular engagement with companies unwilling or unable to comply with the government’s procurement regulations. While the Energy Policy Act of 2005 granted OTA to DOE at the agency level, it did not authorize the labs to use it directly. OTA could be a more effective model for technology transition if driven at the local laboratory level, where the interaction with industry is vital for success. 
\end{quote}

\begin{itemize}
    \item {\bf Industry and national laboratories should develop public-private partnerships to accelerate scientific discovery and benefit industry applications, particularly in emerging technologies. (CEF01 Recommendation 1)}
\end{itemize}

\begin{quote}
    Promoting public-private partnerships will accelerate HEP innovations. In specific technology areas such as accelerators, US federal program managers proposed developing public-private partnerships to foster and support small and large technology businesses who collaborate with the laboratories and serve as commercialization partners for critical technologies developed as part of facilities and experiments in high-energy physics. These public-private partnerships could serve as both advocacy and economic development entities for HEP-derived technologies, as well as matchmakers that aid companies and laboratories in forming collaborations, which lead to positive commercialization outcomes.
    
    Entrepreneurial Leave Programs (ELP) allow employees to take a leave of absence or separation from the laboratory in order to start or join a new private company. ELPs encourage startup activities by reducing the risks faced by the employee entrepreneur. Some elements of an ELP may include business preparation/training, a means for licensing laboratory IP, continuity of health benefits during leave, and a mechanism for returning to work. ELPs are not implemented consistently across the DOE complex; some laboratories have ELPs while others do not.
    
    Providing more technology transfer educational opportunities targeted to HEP researchers will facilitate a ramp-up to I-Corps level of engagement for high-energy physics researchers. This would be a great opportunity to provide researchers the building blocks to enable more engagement in capturing innovations. Discussions on the types of intellectual property (ex. patents, copyrights), rights afforded to researchers from their innovations, and the mechanisms to engage with industry to advance their technologies would provide valuable resources and new perspectives to HEP personnel.
\end{quote}

\subsection{Technology transfer with Scaleups}
HEP collaboration with small businesses is primarily facilitated through SBIR program Phase I awards, which are relatively plentiful I and easy to access. However sustaining development through this channel is difficult because Phase II awards are more limited in number and lack sufficient funding for deep tech transfer. Work with large companies through Cooperative Research and Development Agreements (CRADA) is a slow and lengthy process. In the long term, CRADA partnerships are quite fruitful, but the slow starts and small number of opportunities create a high barrier for execution.

A relatively unexploited and overlooked intermediate option is partnership with mid-sized scaleup companies in the post-startup phase. These middle-ground businesses can be identified using a bottom-up approach of scouring online databases such as crunchbase.com and dealroom.com. A top-down approach of building relationships with venture capitalists (VC) with vested interests in scaleup firms.

\begin{itemize}
    \item {\bf HEP should leverage multiple programs and relationships to build collaborations with scaleup companies on HEP projects.}
\end{itemize}

\begin{quote}
    Laboratories should host “Discovery Days” for scaleups. National laboratory business development (BD) efforts with larger companies involve hosting Discovery Days. Product leads and problem owners from companies are invited to visit the lab and have discussions with technical/domain experts. Success is measured in the number of Discovery Days that convert to project collaborations. Similar Discovery Days to host scaleups at the lab to deep dive into their technology roadmap need to be established.
    
    Venture Capital firms generally prefer to procure services from a commercial service provider instead of a lab, partly because of the relative difficulty getting technology out of the lab due to how hard it is to get an exclusive license. Labs find it better to give exclusive license to large corporations, as opposed to startups where there is more chance the company might fail. Leveraging contacts at venture capital firms is extremely useful, since they have access to service providers such as lawyers, accountants, government lobbyists, and labs. The labs create a summary description of what they do (i.e. their value proposition), which VCs can then share with companies to see if they are interested. 
    
    HEP should also leverage existing university relationships to connect with Venture Capitalists. Some VCs such as ARCH look for strong scientific founders who can help their companies at early stages including getting an exclusive license from the university, which is not as easy to get from a national lab. The university model, which allows staff members to spend one day a week on external projects, helps facilitate such work. 
\end{quote}

\subsection{Application-driven engagement with universities, synergies with other funding agencies}
Laboratory-university HEP partnerships have been very successful, but within the United States have been limited to university physics departments. However, as technology advances, the level of engineering design for accelerators and detectors keeps growing and close collaboration between the labs and university engineering departments is becoming more important. Interactions between the labs and engineering departments are opportunistic and transactional rather than systematic and synergistic like with physics departments. In contrast to Europe, the US HEP community does not have the programmatic ability to directly support application engineering research in the academy. In areas such as computation and microelectronics, upcoming HEP technology projects will require significant amounts of engineering R\&D. This activity and those conducting it need to be considered integral to the HEP community. In addition, HEP labs need to partner with universities to produce a technology workforce.

\begin{itemize}
    \item {\bf Funding agencies should work with universities to create cross-agency engineering initiatives focused on application-driven fundamental technology rather than fundamental science. (CEF01 Recommendation 3)}
\end{itemize}

\begin{quote}
    The DOE Office of HEP should engage more with engineering departments to create explicit representation of engineering partnerships in established listings of funding opportunities.  Clearly label support for HEP-Engineering efforts, such as fellowships reserved for engineering graduate students, or a class of projects designated as science-engineering partnerships, where the project application requires dedicated components reserved for both science and engineering (similar to NSF grant funding where a research and an education component are requested to be described individually). These engineering collaborations should be expanded across DOE and with other agencies (e.g. NASA, NSF, DOD).
\end{quote}

\begin{itemize}
    \item {\bf HEP labs and universities should work together to develop intentional pipelines of engineers for the HEP workforce from undergraduate students to professional ranks.}
\end{itemize}

\begin{quote}
    A possible pipeline could be as follows. HEP labs recruit undergrads for internships through partner universities $\rightarrow$ students are trained in the setups and topics the HEP lab prioritizes $\rightarrow$ successful undergraduate students are channeled to graduate programs across a network of partner universities $\rightarrow$ universities recruit these students into engineering PhD programs $\rightarrow$ co-advising models are used to mentor these students by both an engineering professor and a HEP scientist $\rightarrow$ feed the students back into the HEP workforce.
    
    In addition, universities should welcome input from HEP labs on recruiting the next generation of graduate students from the cohort of international students with interdisciplinary (physics, science, and engineering) backgrounds. Schools could also establish joint academic appointments for HEP lab and industry scientists within engineering departments. These HEP scientists could then also be thesis advisors and thesis committee members of students.
    
    HEP labs should promote engineering students for awards, such as the URA Visiting Scholar award. Often times, engineering faculty and engineering PhD students are not aware of all opportunities that exist within the HEP and national lab ecosystems. Guidance from HEP scientists will help lift entry barriers for them.
\end{quote}

\subsection{Big Industry Engagement to Benefit HEP: Microelectronics Support from Large CAD Companies}
Only a few large companies have both deep expertise in modern microelectronics and access to CAD-EDA tools. In addition, ASICS designed for the extreme environment requirements characteristic of HEP have little market value for those large companies. Therefore, microelectronics for HEP tend to be designed by partnerships of national lab and university personnel. However, these collaborations typically do not have access to the suite of CAD tools due to complicated and expensive licensing frameworks, which are negotiated independently by each DOE lab. The DOE needs to develop a centralized licensing framework with CAD vendors to bring economy of scale and flexibility for each lab to procure the set of tools required for their own projects and teams. There are motivating benefits to the microelectronic industry to pursue this business relationship with HEP. Extreme environment microelectronics make up little of the commercial market, but that segment is growing, particularly for QI and AI applications. DOE science users are typically good sources of feedback on cutting edge uses of advanced CAD-EAD tools, and also develop the talent pool for the microelectronics workforce. The following recommendations arose from DOE HEP hosted meetings with several major CAD-EAD companies.

\begin{itemize}
    \item {\bf A collective all-of-DOE approach for engaging Big Industry should be employed for procurement of common industry tools, licenses and services. (CEF01 Recommendation 5)}
\end{itemize}

\begin{quote}
    Setting DOE-wide common terms and conditions with the flexibility for each lab to make the technical choices specific to their program and negotiating low-cost research licenses for basic science developments will enhance project collaboration with big companies. DOE should consider creating a Collaborative Innovation Hub scoped for cooperative team shared access to CAD/EDA tools, training, and support. Establishing dedicated cloud-based communal participation platform between academia, DOE national labs, and CAD/EDA companies, and leveraging successful solution frameworks (e.g. DARPA Innovation Package, Europractice IC Service, DOD Cloud Access Rights) will bring efficiencies of shared access.
    
    It will also be useful to Incorporate some aspects of CAD/EDA companies’ academia policies for research projects at national labs, in order to create a new class of research licenses. The resulting solutions should keep intact the premise of CAD/EDA companies’ contributions with special arrangements for commercializing research results. The academic network can also be leveraged to cultivate the talent to advance and promote innovations in semiconductor technologies.
\end{quote}

\subsection{Transformative Technology for FLASH Radiation Therapy}
Radiation therapy (RT) cancer treatment has arguably delivered the greatest societal impact of any particle accelerator application, and a large share of accelerator science has been enabled through HEP research support. FLASH radiation therapy (FLASH-RT) is a recent development in which ultra-high doses of therapeutic radiation are delivered in less than a second. Experiments show that FLASH-RT effectively destroying tumors while almost completely sparing normal tissue. However, there are technical difficulties in the development of a clinically-safe delivery system, and the accelerator capabilities within the HEP community will be needed to resolve the issues.

Most R\&D for FLASH has been carried out using 4-6 MeV electrons from clinical linacs, producing strong results. Photon beam FLASH studies using synchrotron radiation and X-ray tubes have yielded mixed results. Some work has also been done with 230-250 MeV shoot-through beam protons from CW and iso-cyclotrons. Limitations include intensity requirements preventing the use of energy degraders for proton beams, and synchrotrons lacking the intensity of ion beams required for FLASH. 

\begin{itemize}
    \item {\bf Prioritize and simplify high risk, high reward transformative technology opportunities. (CEF01 Recommendation 6)}
\end{itemize}

\begin{quote}
    In some technical areas (e.g. FLASH radiotherapy), high impact technology incubation by the HEP ecosystem can produce significant, and occasionally disruptive, benefits to society, within a decade timeframe. In these scenarios, we recommend prioritizing and simpliying access by all domestic stakeholders to HEP facilities, expertise, and resources. 
    
    The HEP community should carry out a broad R\&D program to clinically realize the curative potential of FLASH-RT with different radiation modalities. Among the relevant projects are:

    \begin{itemize}
        \item The Advanced Compact Carbon Ion Linac (ACCIL) is a program initiated by the Argonne National Laboratory to develop up to 1 kHz repetition rate, compact proton linac capable to deliver FLASH-RT doses.;
        \item Scaling Fixed Field Gradient Accelerators (FFGA) are synchro-cyclotron style proton accelerators, which can operate at high repetition rates and high currents consistent with FLASH needs; most of the current R\&D programs on scaling FFGAs are performed by Japanese research groups.;
        \item Non-scaling FFGAs are particularly well suited for accelerating other ion species (i.e. carbon), and there is a pilot facility under construction at the National Particle Beam Therapy Center (Waco, TX).;
        \item Laser-driven accelerators can deliver very large doses of protons or high energy electrons from a compact source (both scenarios are potentially of interest to FLASH-RT). The bulk of US program is centered at the LBNL BELLA laboratory.;
        \item The pulsed power based linear induction accelerator (LIA) using a multilayered bremsstrahlung conversion target also represent very promising technology in meeting FLASH-RT requirements, and there is a pilot program underway at LLNL.;
        \item Multiple groups are also working to develop FLASH-capable X-ray systems, including the ROAD initiative by UCLA/RadiaBeam, and PHASER initiative by SLAC/Tibaray.;
        \item One potential application, which can take advantage of the recent interest by HEP community towards novel cold RF technology, is a compact cold-RF Very High Energy Electron (VHEE) radiotherapy system, with relevant R\&D programs initiated at SLAC and at CERN.;
    \end{itemize}    
\end{quote}

\subsection{Nurturing the Industrial Accelerator Technology Base in the US}
It is widely perceived by the HEP accelerator community that accelerator technology transfer to US industry is not a high priority. US HEP commonly develops state-of-the-art accelerator technology, then buys it back later from international firms for domestic projects. Europe and Asia have nurtured vibrant accelerator industrial bases, leaving US firms at a competitive disadvantage. This has resulted in a US accelerator community plagued by increased costs, insufficient component availability, dependence on foreign sources, small talent pool of technical personnel and low societal recognition of accelerator science benefits.

Although industrial firms play critical roles in the scientific enterprise, important US accelerator companies have struggled to survive. Pioneers in SCRF and undulator technology enjoyed initial success, but failed due to the inability to sustain support from DOE research, leading to the loss of capital and unique expertise. Regulatory and policy recommendations are suggested to build a competitive domestic industrial base for accelerator technology.

\begin{itemize}
    \item {\bf DOE should invest in programs to provide direct support to specific critical need industries. (CEF01 Recommendation 7) }
\end{itemize}

\begin{quote}
    There is a growing interest in the community to improve support to the domestic industrial vendors providing critical technological capabilities to the HEP ecosystem. We recommend that DOE takes a proactive approach in establishing critical technology needs, and work directly with the qualified vendors to maintain and develop critical industrial capabilities, relevant to these needs.
    
    Modify the US Small Business Innovation Research (SBIR/STTR) program to nurture these small businesses across the “Valley of Death”. One improvement would be to more closely align the program technical topics to the future procurement needs of the labs, and encourage the labs benefiting from the SBIR funded work to maintain the momentum and work with the industry beyond the SBIR funded phase. DOE should also establish a method to identify key technologies that will be needed in a decade time frame and create new channels of direct funding to the qualified industrial enterprises to develop expertise, infrastructure, and capacity to meet such needs. It is also equally important to be able to help sustain the companies that have already achieved critical capabilities.
    
    Support specialized industrial vendors by implementing directed “knowledge transfer” programs. Recent decades saw a proliferation of national laboratories-based commercialization centers built around the technology transfer activities. Yet, few of them can report success and the idea of technology transfer through funding the commercialization activities by the labs is generally counterproductive for the purposes of building the industrial vendors base. We believe it would be more beneficial to deemphasize technology transfer as a means of supporting the labs, and emphasize knowledge transfer as a means of supporting motivated businesses to expand capabilities of interest to the DOE programs. 
    
    Laboratories should also simplify some of the procurement practices, and likewise explore various creative ways for industry and laboratories to collaborate on the prototype developments that would minimize the risks and maximize return to both sides. The accelerator community should promote programs that facilitate direct and open communication channels between laboratory engineering and technical staff with their industrial counterparts (there are many conferences for scientists to attend and share their experiences, but not so many venues are available to technicians and engineers whose skills are essential and irreplaceable in our field).
\end{quote}

\section{CEF02: Career Pipeline and Development}
\label{sec:CEF02}

\renewcommand{\cefgroup}{2}

This working group is not simply about making early career scientists aware of different opportunities, but also changing the culture of HEP career paths.  It aims to identify and encourage career opportunities for high energy physicists in both academia and industry, and to identify useful partnership options between HEP and industry. Smoother pipelines between different types of employment are critical for the success of HEP trainees in the future. One objective is  to promote the skill development of physics graduates and young researchers and encourage career direction-based scientific majors and skills.

Thirty-two LOI were submitted to this working group and were condensed into three contributed paper topics, namely:
\begin{enumerate}
    \item Facilitating Non-HEP Career Transition;
    \item Enhancing HEP research in predominantly undergraduate institutions and community colleges;
    \item Tackling diversity and inclusivness in HEP.
\end{enumerate}
Topic (3) was integrated to and developed in the topical group on diversity, equity and inclusion in Section~\ref{sec:CEF03}. Ultimately, two contributed papers on topics (1) and (2) were prepared and presented in Refs.~\cite{https://doi.org/10.48550/arxiv.2203.11665, https://doi.org/10.48550/arxiv.2203.11662}.  Considering that there are fewer academic positions than job seekers, many degree holders will eventually seek jobs outside HEP, where, in sectors such as industry, there are demands for skills acquired in HEP training, e.g. data science or machine learning. However, organized guidance---developed through engagements between the HEP community and the alumni that have already transited out of HEP---is needed to help with non-HEP career transitions~\cite{https://doi.org/10.48550/arxiv.2203.11665}. Another career trajectory may to employments at predominantly under-graduate institutions (PUI) and community colleges (CC), with high teaching loads and lack of support for research. PUI and CC can serve as pipelines to improve diversity and under-representation in HEP, by facilitating participation of faculties and students at PUI and CC in HEP activities~\cite{https://doi.org/10.48550/arxiv.2203.11662}. 

\subsection{Facilitating Non-HEP Career Transitions}

It is noted in Section~2 of Ref.~\cite{https://doi.org/10.48550/arxiv.2203.11665} that more that two-third of trained physicists will eventually transitions to employments in private or government sectors, collectively referred to as "industry"; moves to the industry sector may occur at various stages of HEP career evolution and proper planning is needed to facilitate the transition. A survey conducted by the Snowmass Early Career (SEC) physicists included questions about career pipeline and development~\cite{https://doi.org/10.48550/arxiv.2203.07328}, to offer insights on existing efforts, support, networking, preparation and attitude towards career change and alumni participation or eventual return to HEP. Details about the SEC survey and findings related to career pipelines and development are documented in Refs.~\cite{https://doi.org/10.48550/arxiv.2203.07328, https://doi.org/10.48550/arxiv.2203.11665}. We recall here the suggestions:
\begin{itemize}
\item Supervisors and mentors should be directly involved in planning the career of their mentees early on. This career plan should not be based on the desires of the mentor but the skills and interest of the mentee. A commensurate effort in the job search process is also needed.
    
\item Supervisors should allow a certain fraction of working time for their mentees to pursue opportunities and preparation activities for a possible industry career.

\item HEP experiments, laboratories, or university departments should provide training for supervisors so that they can better understand and be more sensitive to the needs of their mentees in terms of their career goals and preparation.

\item HEP experiments and/or laboratories should provide workshops on industry job preparation:  translating HEP skills and examples to industry language, converting CVs to resumes suitable for different fields, finding successful job search phrases (for example, ``Engineer" or ``Data Scientist" as opposed to ``Physicist").  This will be most successful when paired with the recommendations below for deepening connections with HEP alumni.
    
\item HEP experiments and/or laboratories should develop innovative opportunities for networking with HEP alumni in various fields to strengthen industry job search success. Alumni are more than willing and happy to respond and engage. This will be most successful when paired with the recommendations below for deepening connections with HEP alumni.
\end{itemize}

The survey revealed that about 50\% of respondents tried to find jobs in their field before moving on to opportunities in industry; they exited at student or post-doctoral levels and went to STEM-related responsibilities outside academia.  These transitions were mostly facilitated through networking; but the difficult step is to leave at a relatively late stage in academic. It is challenging to return to HEP but alumni are open to joint projects, and this can help strengthen partnerships between HEP and industry~\cite{https://doi.org/10.48550/arxiv.2203.11665}:
\begin{itemize}
\item Supervisors and mentors should actively communicate with alumni and highlight their experiences for current students and postdocs, to normalize the reality of transitioning to an industry career.

\item The US HEP community should develop tools and portals for connecting with alumni. Existing programs for networking with alumni like at CERN must be studied and adapted. This effort should be supported and strengthened by funding agencies by dedicating a small amount of continuous funding to support technical and personnel staff that can organise and build a framework that can serve as a hub to facilitate process of networking with alumni. A DOE lab would be an ideal place to host this effort, like Fermilab, which is a hub for US particle physics.
\end{itemize}

HEP experiments and laboratories should take creative steps to reverse ``brain drain'' from HEP by exploring mechanisms for collaboration with alumni on HEP projects:
    \begin{itemize}
        \item Alumni are a relatively low cost but very valuable asset with an abundance of experience from transitioning to an industry career. Their goodwill to contribute and strengthen ties with HEP can be tapped to facilitate industry job transitions and further the goals of both groups.
        \item Individual scientific collaboration can be extended to the company of the alumni itself and this can strengthen knowledge transfer from labs and universities and vice versa; and work done by HEP research can benefit companies and vice versa.
    \end{itemize} 

HEP training offered at universities and laboratories could be extended to industry for opportunities to apply HEP skills in a different environment and culture and this may facilitate eventual career transitions. The survey showed support to develop such HEP-industry partnerships~\cite{https://doi.org/10.48550/arxiv.2203.11665}:

\begin{itemize}
\item HEP laboratories should create targeted internships or training programs in the areas of Accelerator Technology, Computer and Information Science, Detector and Engineering Technology, Environmental Safety and Health and Radiation Therapies. This would expand access to industry-focused training to students and postdocs who are not based at national laboratories.
    
\item HEP laboratories should leverage existing public-private partnerships with industries like Accelerator Technologies, Computers Information Science, Detector and Engineering Technologies and also Environmental Safety to create experience for resident students and early career scientists to build skills ad connections for a future industry career.

\item Funding agencies should evaluate funding rules and regulations to allow HEP students and postdocs to pursue industry-focused training that can be integrated with their core research curriculum.
    
\item Supervisors must adopt a mindset that industry partnerships and career transitions are valuable options for their students and postdocs, and should support their participation in training opportunities whenever possible.
\end{itemize}

\subsection{Enhancing HEP research in predominantly under-graduate institutions and community colleges}

HEP activities are carried out primarily by people at laboratories and research focused non-PUI. However, about 40\% of undergraduate students in the United States are enrolled in CC where $\sim$80\% are from demographics under-represented in STEM. It is therefore important for the HEP to engage the vast community at PUI and CC which may serve as pipelines to improve diversity and under-representation in HEP.  For such engagements to be productive, barriers to participation of PUI and CC faculties in HEP activities must be addressed. These barriers include heavy teaching loads, lack of guidance and research funds, lack of research infrastructure and equipment, and lack of administrative support and understanding of the regulations and requirements for successful participation in HEP---see Ref.~\cite{https://doi.org/10.48550/arxiv.2203.11662} and the references therein. To address these barriers, we suggest the set of recommendations on institutional culture~\cite{https://doi.org/10.48550/arxiv.2209.10114}:

The HEP community should encourage a global shift in perception, acknowledging that:
    \begin{itemize}
        \item Undergraduate research experiences are {\it key} to engaging a broader section of the student population.
        \item PUI or CC faculties have much to offer their collaborations, particularly in experiment-wide training and educational activities.
    \end{itemize}

HEP experiments should offer coordinated communication from leadership to PUI administrators, extolling the features of high energy physics research alongside highlighted participation.

The HEP community should offer special sessions for PUI and CC faculty at national meetings to develop a deeper sense of community.

 We also have suggestions for research funding~\cite{https://doi.org/10.48550/arxiv.2209.10114}:  

\begin{itemize}
\item Funding agencies should strengthen participation by PUIs in HEP by allocating funds for grants from these institutions, and \textbf{HEP experiments or laboratories} should fund grant-writing workshops. 

\item Funding agencies should allow course buyouts in proposals by PUI/CC faculty in order to boost productivity and establish continuity in PUI research programs.
        
\item Funding agencies, HEP experiments, and laboratories should create or support paid summer programs for PUI faculty to work at National Labs or non-PUIs, as well as research opportunities for students not enrolled at major HEP institutions.
    
\item Supervisors and HEP experiments should provide training to interested students and postdocs on US-specific research funding procedures. 
\end{itemize}

 Finally, We make the following suggestions for participation in HEP activities~\cite{https://doi.org/10.48550/arxiv.2209.10114}:  

\begin{itemize}
\item Non-PUI senior-level researchers should investigate how their groups could offer opportunities for short-term and long-term collaboration on their experiment to faculty and/or students at local PUIs.

\item HEP experiments must reevaluate large fixed ``entry fees" per institution, if they exist. Consider implementing ``light" membership forms that are low cost but not time limited.
    
\item US HEP experiment leaders should advocate with international experiment leadership for pathways to sustainable membership for PUIs, which are most common in the US. Postdocs should be aware of options for entering these pathways so they are not discouraged from applying to PUI faculty positions.
    
\item HEP experiments must continue to improve options for remote participation in experiment meetings and service tasks, especially operational shift work.
\end{itemize}

\subsection{Connections with other Frontiers and Topical Groups}

Improvement in career pipeline and development requires improvement physics education as discussed in Section~\ref{sec:CEF04} to prepares a skilled workforce needed for HEP and career migrations industry and improve diversity, under-representation and inclusion in HEP as discussed in Section~\ref{sec:CEF03}. HEP experimental physicists have developed expertise and transitioned into accelerator physics; this is essential to HEP operations and applications in industry, such as medical, materials, pharmaceutical, chemical and biological areas. Small scale experiments in neutrino, dark matter, nuclear and rare processes are sources of training for HEP physicists and transitions to industry.  Technology transfers and applications and industry as discussed in Section~\ref{sec:CEF01}, in addition to training in instrumentation and detector technologies,  can enhance fruitful career transitions~\cite{https://doi.org/10.48550/arxiv.2209.10114}.

\section{CEF03: Diversity, Equity and Inclusion}
\label{sec:CEF03}

\renewcommand{\cefgroup}{3}

This topical group focused on issues and projects related to (1) Diversity, (2) Inclusion, (3) Equity and (4) Accessibility; all are essential not only to professional success in our field, but to developing a better society at large. The group gathered information concerning diversity, equity, inclusion and accessibility in high energy physics, instances of success and failure, and actions that have been taken to promote our tenets. Thirty-two letters of interest were tagged to this group; other inputs came from surveys, town hall meetings and discussions. Ultimately, twelve contributed papers were developed, as detailed in Ref.~\cite{https://doi.org/10.48550/arxiv.2209.12377}. The contributed papers may be categorized as follow:
\begin{itemize}
    \item Accessibility in High Energy Physics: Lessons from the Snowmass~\cite{AccessibilityInHEP};
    \item Lifestyle and personal wellness in particle physics research~\cite{LifestyleAndPersonalWellness};
    \item Climate of the Field: Snowmass 2021~\cite{ClimateOfTheField};
    \item Why should the United States care about high energy physics in Africa and Latin America~\cite{HEPInAfricaAndLatinAmerica};
    \item Experiences of Marginalized Communities in HEP~\cite{MarginalizedCommunities, HowToReadSelectedSnowmassPapers, PowerDynamicsInPhysics, PolicingAndGatekeepingInSTEM, InformalSocializationInPhysicsTraining};
    \item In Search of Excellence and Equity in Physics~\cite{ExcellenceAndEquityInPhysics};
    \item Strategies in Education, Outreach, and Inclusion to Enhance the US Workforce in Accelerator Science and Engineering~\cite{StrategiesToEnhanceAcceleratorWorkforce}.
\end{itemize}

\subsection{Accessibility in High Energy Physics: Lessons from the Snowmass Process}
Various barriers may impede on full participation in HEP activities; in Ref.~\cite{AccessibilityInHEP}, using the results of surveys, experiences and additional feedback from community members, and best-practice guidelines, the authors studied accessibility to engagements in HEP and offered recommendations for improvement. The authors discussed the resources and funding needed to implement the recommendations. Barriers to accessibility include lack of financial support, mental health issues, deaf/hard of hearing, visual disability/blind, caretaker responsibilities and virtual access. These barriers affect the community as a whole by impacting on the ability to collaborate with the members that face accessibility challenges. Survey respondents said that logistics for accessibility should not be the burden of the persons that need access; the availability of transcripts and auto-captioning were noted; however, these fail to transcribe correctly all ramifications of human expressions. Furthermore, resources available often require advance planning and funding and these are the  core recommendations for organizing accessible physics events. More details on the studies done and recommendations can be found in Ref.~\cite{AccessibilityInHEP}. 

\subsection{Lifestyle and personal wellness in particle physics research activities}
The demand of particle physics activities may result in an unhealthy imbalance between work and personal life. Unequal remunerations, living conditions and caretaker responsibilities and competitiveness for career progression, visibility, and grants, are among the causes for poor work-life balance. These career requirements lead to working after-hours, during weekends and holidays; teleworking may be impacted by living conditions and may blur the boundaries of work and personal times. Such an imbalance may result in mental health issues, burnouts, poor job satisfaction, and poor performance~\cite{LifestyleAndPersonalWellness}. Other triggers of work-life imbalance are the expected activities that do not translate into research outputs or are not compensated in career evaluations---these include work for community engagement as noted in Section~\ref{sec:CEF05}, for DEI initiatives, mentorship, refereeing, reviews, hiring committees, etc. Furthermore, as noted Refs.~\cite{HowToReadSelectedSnowmassPapers, PowerDynamicsInPhysics, PolicingAndGatekeepingInSTEM, InformalSocializationInPhysicsTraining}, unwelcoming working environment that translates into discrimination, harassment, non-inclusion, code of conduct violations, etc., places undue burden on the victims and members of marginalized communities and lead to work-life imbalance. In Ref.~\cite{LifestyleAndPersonalWellness}, the authors propose recommendations or actions to improve work-life balance.

\subsection{Climate of the Field: Snowmass 2021}
The state of existing policies and their effectiveness to create an inclusive, equitable and safe environment for HEP engagements are discussed in Ref.~\cite{ClimateOfTheField}---``climate of the field''. In many scientific engagements, code of conduct guidelines are in place to define respectful interactions. Mechanisms to address violations are also defined. Yes, implementation of these guidelines and how violations are reported and addressed, are affected by the ``climate of the field''. An example is how a violation is handled when it occurs in a collaboration and the concerned parties (perpetrator and victim) have different institutional affiliations. Often, group dynamics and power dynamics lead to an inability to adhere to code of conduct guidelines and address violations; this creates an unwelcoming environment and alienates victims and marginalized folks; as noted in Ref.~\cite{LifestyleAndPersonalWellness}, it also impacts on work-life balance. The contributed paper of Ref.~\cite{ClimateOfTheField} provides recommendations for several top-down approaches that should be implemented by the community as well as recommendations for funding agencies to support these approaches.

\subsection{Why should the United States care about high energy physics in Africa and Latin America?}
Contributions of developing countries to high energy physics activities are hampered by limited resources and national priorities. Title VI of the 1965 Higher Education Act~\cite{TitleVI}, designed "to support US national interests and maintain global competitive edge in the international arena", is a compelling reason for the United States to support HEP in developing countries. Mechanisms and recommendations to improve HEP engagements with Africa and Latin are articulated in Ref.~\cite{HEPInAfricaAndLatinAmerica} where it is argued that such sustained engagements will help international development, improve diversity and increase the participation of developing countries in HEP.

\subsection{Experiences of Marginalized Communities in HEP}
\label{sec:obj}

Power dynamics, informal socialization, policing and gate keeping in HEP create an environment and culture that alienate under-represented physicists and negatively affect their participation. Often, privileged folks lack awareness and attention or focus on perception rather than reality; therefore, they hang on to claims of objectivity in physics which only serve to maintain the culture of under-representation and non-inclusion, to deny the negative experiences of marginalized physicists, and to expect them to shrug off these bad experiences, despite the harm caused, in order to be taken seriously as physicists. The key to improve the experiences of BIPOC physicists consists of addressing the complexity and impact of power dynamics, policing and gate keeping~\cite{PowerDynamicsInPhysics, PolicingAndGatekeepingInSTEM}, and implementing actions individuals and organizations can take to lower barriers for early career BIPOC physicists~\cite{InformalSocializationInPhysicsTraining}.

Despite all the efforts and investment to improve DEI in HEP, these issues remain as demonstrated in Refs.~\cite{HowToReadSelectedSnowmassPapers, PowerDynamicsInPhysics, PolicingAndGatekeepingInSTEM, InformalSocializationInPhysicsTraining, MarginalizedCommunities}.  We offer concrete suggestions towards improving DEI; these suggestions include effective approaches to reach members of marginalized communities through engagements as discussed further in Section~\ref{sec:CEF05} and Ref.~\cite{MarginalizedCommunities}.

\subsection{In Search of Excellence and Equity in Physics}
The claims of objectivity in physics lead to meritocracy, i.e. the idea that ``scientific work is judged on its merits and that opportunities in physics are equitably available to all aspirants''. However, as demonstrated in Ref.~\cite{ExcellenceAndEquityInPhysics}, there is far more under-representation than could be expected from meritocracy. This further challenges the claims of  objectivity in physics, along similar lines as Refs.~\cite{HowToReadSelectedSnowmassPapers, PowerDynamicsInPhysics, PolicingAndGatekeepingInSTEM, InformalSocializationInPhysicsTraining, MarginalizedCommunities}. To address this, changes in community practices are needed; we should challenge or verify organizational claims of equitable access. Focused efforts, continuous measurements and frequent corrections are required to achieve fair procedures, eliminate barriers and improve under-representation~\cite{ExcellenceAndEquityInPhysics}. The physics community should seek best practices on how to combine equity and excellence; the private sector has made some progress to match best practices to company values, thus mitigating damages resulting from public exposure of misconduct. The need to enforce codes of conduct and to address violations, mentioned in Ref.\cite{ClimateOfTheField}, are also echoed in Ref.~\cite{ExcellenceAndEquityInPhysics}. The American Physical Society has made efforts towards an  equitable, diverse and inclusive field; however, community participation is required to improve excellence and equity. Meritocracy, instead of cronyism, is important to identify leaders. In Ref.~\cite{ExcellenceAndEquityInPhysics}, the authors go further to suggest recommendations towards an ethical hiring process for excellent leaders that will uphold the values of equity.

\subsection{Strategies in Education, Outreach, and Inclusion to Enhance the US Workforce in Accelerator Science and Engineering}
Accelerators, large or small, play important roles in fundamental research and applications; they are essential to discovery science and high technology, thus can help to train a strong technical workforce needed for particle physics research. In Ref.~\cite{StrategiesToEnhanceAcceleratorWorkforce}, the educational and outreach opportunities available in accelerator science and engineering are reviewed, with the objectives to attract talents and develop capacity for future R\&D; in this process, the need to improve diversity, equity and inclusion is noted---the participation of women in the US Particle Accelerator School (USPAS) has increased; however, under-representation of women and historically marginalized groups still persists. Recommendations are proposed to improve diversity, equity and inclusion in accelerator science and engineering~\cite{StrategiesToEnhanceAcceleratorWorkforce}.

\subsection{Suggestions to Improve Diversity, Equity and Inclusion in HEP}
\label{sec:recommendations}

From the aforementioned work in Section~\ref{sec:CEF03}, we have prepared suggestions and resources that are tailored to particle physics, cosmology, and astrophysics, to further promote diversity and encourage equity, inclusion and accessibility at all levels of scientific discourse, engagements and managements.

\subsubsection{Suggestions for Funding Agencies}

HEP communities should improve use of robust strategic planning procedures, including a full re-envisioning of science workplace norms and culture:

\begin{itemize}
    \item Prioritize community-related issues at the funding level, e.g. inclusion of community-related topics into safety parts of collaboration ``Operational Readiness Reviews," ``Conceptual Design Reviews," or similar documentation submitted to funding agencies. Funding agencies should provide clear and enforceable requirements for the advancement of DEI issues in grants, programs, and evaluations~\cite{ClimateOfTheField}.
    
    \item Funding agencies should provide formal recommendations for institutions, research groups and collaborations for handling violations of their codes of conduct. This should include advice on handling community threats, removal of collaboration affiliates, leadership rights and responsibilities, and protections against legal liability for leadership that is responsible for that enforcement. This should also include advice on reporting to the funding agency itself; if there is no mechanism for reporting misconduct to a funding agency, that mechanism should be developed~\cite{ClimateOfTheField}.
    
    \item Funding and structural aid should be made available to develop ``Collaboration services'' offices at host laboratories. Such offices should provide HEP collaborations and other physics communities of practice with the following: a) advice on legal and policy topics, b) training in project management and ombudsperson training, c) logistical tools including facilitation of victim-centered investigation and mediation, d) resources and funding for local meeting accommodations, and other topics as described here in Section \ref{sec:recommendations}~\cite{ClimateOfTheField,AccessibilityInHEP, InformalSocializationInPhysicsTraining, PowerDynamicsInPhysics, ExcellenceAndEquityInPhysics}.
    
    \item Funding agencies should use their leverage to promote community-focused policies at funded institutions. Funding agencies should require institutions that receive funding to implement policies on vacation time, parental \& family leave (for all genders), and health leave for all levels. Funding agencies should require institutions to prohibit confidentiality in settlements for egregious behavior (e.g. harassment); this promotes accountability and prevents known perpetrators from continuing to harm their communities~\cite{LifestyleAndPersonalWellness, ClimateOfTheField}.
    
    \item Funding agencies should establish a dedicated Office of Diversity, Equity, and Inclusion to work with Program Officers to strategize and prioritize funding decisions and develop equitable practices for the review processes~\cite{MarginalizedCommunities}.
\end{itemize}

HEP communities must implement new modes of community organizing and decision-making that promote agency and leadership from all stakeholders within the scientific community:

\begin{itemize}
    \item Funding agencies should facilitate Climate Community Studies. Studies should not be the responsibilities of individual communities. These studies should be informed by expertise in social and organizational dynamics~\cite{ClimateOfTheField}.
    
    \item Grant calls and assessments should include clear definitions of the tasks expected of PIs, including DEI related tasks, and provide grant funding for each. Alternatively, agencies could provide specific grants and awards for EDI and mentorship work. Agencies should ensure that they pay those on their grant review panels for their time~\cite{LifestyleAndPersonalWellness}.
    
    \item Funding agencies should collect, analyze, and publish demographic information on grant proposals and funded grants. PI and funded \emph{and unfunded} researcher demographic information on grant proposals should be collected and used to track the effectiveness of these measures and are necessary to inform any additional policy changes needed to advance DEI policies and structures~\cite{LifestyleAndPersonalWellness, InformalSocializationInPhysicsTraining}. 
    
    \item Pay for fellowship- / grant-funded student and postdoctoral researcher positions must increase. Pay should include cost-of-living adjustments, health / wellness / leave benefits, and relocation expenses~\cite{LifestyleAndPersonalWellness}.
    
    \item The US HEP community should maintain the current engagements and increase investments in Africa and Latin America to improve the reach of HEP in these regions. Funding agencies and international collaborations should acknowledge the disparity in economic capabilities of countries in Africa and Latin America compared to what is available in the United States. Funding agencies should support the development of HEP in these countries, should support and lead initiatives for more equitable contributions (e.g. membership and operations fees for participation in large collaboration, conference fee waivers and travel support to US-based meetings, etc.)~\cite{HEPInAfricaAndLatinAmerica}.
\end{itemize}   

HEP communities should develop partnership with scholars, professionals, and other experts in several disciplines, including but not limited to anti-racism, critical race theory, and social science:

\begin{itemize}
    \item Funding should be made available to both engage with and compensate experts in DEI, anti-racism, critical race theory, and social science. This can take the form of independent grants, but more effective would be the inclusion of climate-related topics into safety components of collaboration ``Operational Readiness Reviews,'' ``Conceptual Design Reviews,'' or similar documentation submitted to funding agencies~\cite{ClimateOfTheField}.
    
    \item Community studies should be run by and receive advice from experts in sociology and organizational psychology. The tools used to evaluate the climate of HEP need to be adequate, effective, and informative. These studies and accompanying expertise should be funded at the federal and institutional levels. They should include evaluation of leadership selection, development of junior scientists and their trajectories, and the existence of detrimental power dynamics that specifically affect underrepresented groups. Undesirable systems should be addressed with direct intervention~\cite{ClimateOfTheField, PowerDynamicsInPhysics, InformalSocializationInPhysicsTraining}.
    
    \item Grant calls and assessments should include the advice of professionals in DEI and education. Such experts should review the entire process, including portfolios in their entirety, but with specific attention to mentorship and DEI plans. Experts should be paid for their time~\cite{LifestyleAndPersonalWellness}.
\end{itemize}

\subsubsection{Suggestions for HEP Communities}

HEP communities should develop or improve robust strategic planning procedures, including a full re-envisioning of science workplace norms and culture:

\begin{itemize}
    \item HEP communities should support and take advantage of existing support structures and informational networks. Tools exist to support efforts to improve diversity and inclusion, as well as to address injustices in our communities. These include the American Association for the Advancement of Science (AAAS) Diversity and the Law program~\cite{AAASDiversityAndTheLaw} which hosts resources to enable promotion of legal and policy goals related to DEI. Knowledge like that collected by the American Institute of Physics’ National Task Force to Elevate African American Representation in Undergraduate Physics \& Astronomy (TEAM-UP) Project~\cite{AIPTeamUpWebsite,TEAMUPreport2020} and the AAAS’s STEMM Equity Achievement (SEA) Change~\cite{AAASSeaChange} should also be promoted~\cite{PowerDynamicsInPhysics}.
    
    \item Institutions and HEP communities must develop reporting mechanisms and sanctions for egregious behavior. These institutions and communities should transparently describe those mechanisms in full for the benefit of all affiliates. Communities must be prepared to exercise those mechanisms. Future HEP community codes of conduct should align with, and current codes of conduct should be reviewed upon new recommendations from funding agencies regarding enforcement and disciplinary measures~\cite{ClimateOfTheField}.
    
    \item The community should prioritize the implementation of best practices networks across institutions and communities of physics practice. This may be facilitated through Collaboration Services Offices, but may also include the facilitation of networks between DEI groups at similar collaborations~\cite{ClimateOfTheField}.
    
    \item All community affiliates should reject harmful rhetoric and behavior related to work-life balance. This includes ``ideas around ‘lone geniuses’, the need for unhealthy work schedules, and the idea that sacrifice of personal wellness demonstrates your commitment to science"~\cite{LifestyleAndPersonalWellness}. Senior scientists are responsible to ensure that they are managing their time and the time of those in their group properly to respect work-life balance (including reducing meetings outside of working hours, or rotating meetings to accommodate varying time zones)~\cite{LifestyleAndPersonalWellness}.

    \item Departments and institutions should have clear definitions of job responsibilities and ensure that they are funding all functions of the job. This includes any DEI work. Assessments should weight work in these areas equally and individuals should be awarded and/or recognized when they excel. Evaluation for employment should be based on carefully developed, public rubrics that include DEI, outreach, and service. Such rubrics should be created with considerable care and research-driven (e.g. if any of the criteria are biased in a way that would limit access or promotion of people who identify with an underrepresented group)~\cite{LifestyleAndPersonalWellness, ClimateOfTheField, InformalSocializationInPhysicsTraining}.
    
    \item Departments and institutions should reject the use of standardized exams in favor of holistic rubrics for admission. Evaluation for admission should reject the use of standardized exams and instead should be based on carefully developed, public rubrics, that are tailored to the department. Such rubrics should be created with considerable care and be research-driven (e.g. if any of the criteria are biased in a way that would limit access or promotion of people who identify with an underrepresented group)~\cite{LifestyleAndPersonalWellness, ClimateOfTheField, InformalSocializationInPhysicsTraining}.
    
    \item Graduate students and postdoctoral researchers should be paid at the level of their respective skill levels. This includes benefits like relocation services, health coverage (including families), retirement savings, subsidized family housing, and are not taxed for fellowship money they do not receive as pay. These benefits must apply while students and their families are abroad on behalf of HEP activities~\cite{LifestyleAndPersonalWellness}.
    
    \item Institutions should have accessible, clear, robust, and flexible policies for parental / family leave (for all genders) and vacation time. These should be guaranteed at all levels. Junior scientists should be made aware of and encouraged to take advantage of institutional policies and resources on diversity, health, leave, vacation, and wellness~\cite{LifestyleAndPersonalWellness}.
\end{itemize}

HEP communities must implement new modes of community organizing and decision-making that promote agency and leadership from all stakeholders within the scientific community:

\begin{itemize}
    \item Reviews of community climate should include an evaluation of how leadership is selected within HEP collaborations, as well as the valuation of sub-community contributions. This should include a expert-advised review of the assignment of high-impact analyses \& theses topics, convenership of working groups, and public-facing roles representing the collaboration such as spokespersons or analysis announcement seminars. Power dynamics within communities should also be evaluated, and should consider the impact that senior scientists can have especially on junior scientists of color. It should also include reviews of the participation of ``non-scientists" in community engagement and authorship, community perceptions of operations and service work, the development of onboarding and early-career networks, and implementation of policies toward equity in information sharing and software~\cite{ClimateOfTheField, PowerDynamicsInPhysics, InformalSocializationInPhysicsTraining, ExcellenceAndEquityInPhysics}.
    
    \item Collaborations should train members in standards in the field and offer mentorship programs to ensure that postdocs and students (especially from underrepresented groups) have additional support and resources. Mentorship programs should be research-driven and should make access to information as ubiquitous as possible. Mentors should help novices navigate the complicated landscape of the community, and care should be taken to address the ``untold rules", like non-academic career trajectories. Information sharing, especially about collaboration policies, procedures, and code-bases, should be evaluated from an equity lens~\cite{LifestyleAndPersonalWellness, InformalSocializationInPhysicsTraining, ClimateOfTheField}.
    
    \item Conferences should offer financial assistance to individuals with hardships. Conferences should offer limited travel grants through an application procedure overseen by an ethics group associated with the conference. To promote the engagement of under-resourced and early-career scientists, conferences should also strongly consider developing an application for sliding-scale / waiver for conference registration fees. Conferences should accommodate care-giving responsibilities by providing childcare onsite, or by supporting the travel of an accompanying person. In both situations, extra funding should be budgeted by the conference to fully or partially cover those costs~\cite{AccessibilityInHEP}.
\end{itemize}

\begin{itemize}
    \item The organizers of all HEP activities should ensure that people with accessibility barriers are truly accommodated, with guaranteed, low-friction, dignified access to all aspects of the experience. All conferences, collaborations, universities, and labs should be made accessible to people with disabilities.  For example, conferences (including virtual meetings) should be announced with enough time to arrange accommodations for any individual needs, and organizers should plan to secure funding and book services far enough ahead of time. Accommodations should include both steno-captioning and ASL interpretation, which should be fully funded as part of the conference budget. Conferences should also be accessible to the blind / low-vision community, which may include screen-reader-accessible tools and ``color-blind-friendly" plots. Other accommodations include: seating or accessible access to amenities like check-in and meals; locating the conference in ADA-compliant buildings with no obstructions to seating, entrances / exits, or accessible pathways; quiet spaces; and designated contacts for troubleshooting accessibility.   An extensive list of recommendations can be found in~\cite{AccessibilityInHEP}.
    
    \item US universities and research labs should encourage and support the participation of their personnel, faculties and research staffs in HEP education and research efforts of African and Latin American countries. US institutes need to partner with Latin America and Africa in establishing bridge programs and supporting community members from Africa and Latin American to come to United States laboratories and universities for research experience programs. Collaborations and conferences should seriously consider decreasing or waiving membership and operations fees for participation and should provide financial assistance for travel to the United States~\cite{HEPInAfricaAndLatinAmerica}.
\end{itemize}

HEP communities must engage in partnership with scholars, professionals, and other experts in several disciplines, including but not limited to anti-racism, critical race theory, and social science:

\begin{itemize}
    \item Experts should be adequately integrated into HEP communities. This is motivated by the need to apply their expertise effectively, and should include collaboration communities. This may take the form of an official collaboration role like a non-voting member of a collaboration council~\cite{ClimateOfTheField}.
    
    \item Identification of leaders within HEP communities should be research-driven. HEP organizations and institutions require leaders who will promote policies and practices that support underrepresented and historically marginalized groups instead of favoring ``politics and convenience". Best practices have been developed by industrial \& organizational psychologists and are under studies at NSF (e.g.\cite{nsfADVANCEOrganizationalChange}). Details on necessary search practices can be found in~\cite{ExcellenceAndEquityInPhysics}. 
\end{itemize}

\subsubsection{Suggestions for Future Snowmass Activities}

\begin{itemize}
    \item Community Engagement topics should be better integrated into other frontiers. This work is the responsibility of \textit{all} HEP community members, and should not be relegated entirely to an independent, volunteer-driven frontier. 
    
    \item Funding for Snowmass activities should include critical infrastructure for accessibility. This includes live captioning for all public events, and infrastructure for hybrid meetings to support those who cannot travel to attend workshops.
\end{itemize}

\section{CEF04: Physics Education}
\label{sec:CEF04}

\renewcommand{\cefgroup}{4}

CEF04, the Physics Education (PE) topical group, examined the role that physics education at all levels plays in advancing the field of HEP. Two goals were identified as critical for the long-term health of the field: 1) attracting students across all demographics to the study of physics, and 2) provide them the education, training, and skills they will need to pursue any career in STEM or related fields. The CEF04 Topical Group Report puts forth recommendations intended to achieve these goals by strengthening ties between researchers and teachers, the academy and the private sector, and domestic and international students~\cite{https://doi.org/10.48550/arxiv.2209.08225}.

\begin{figure}[htb]
\begin{center}
\includegraphics[width=1.0\hsize]{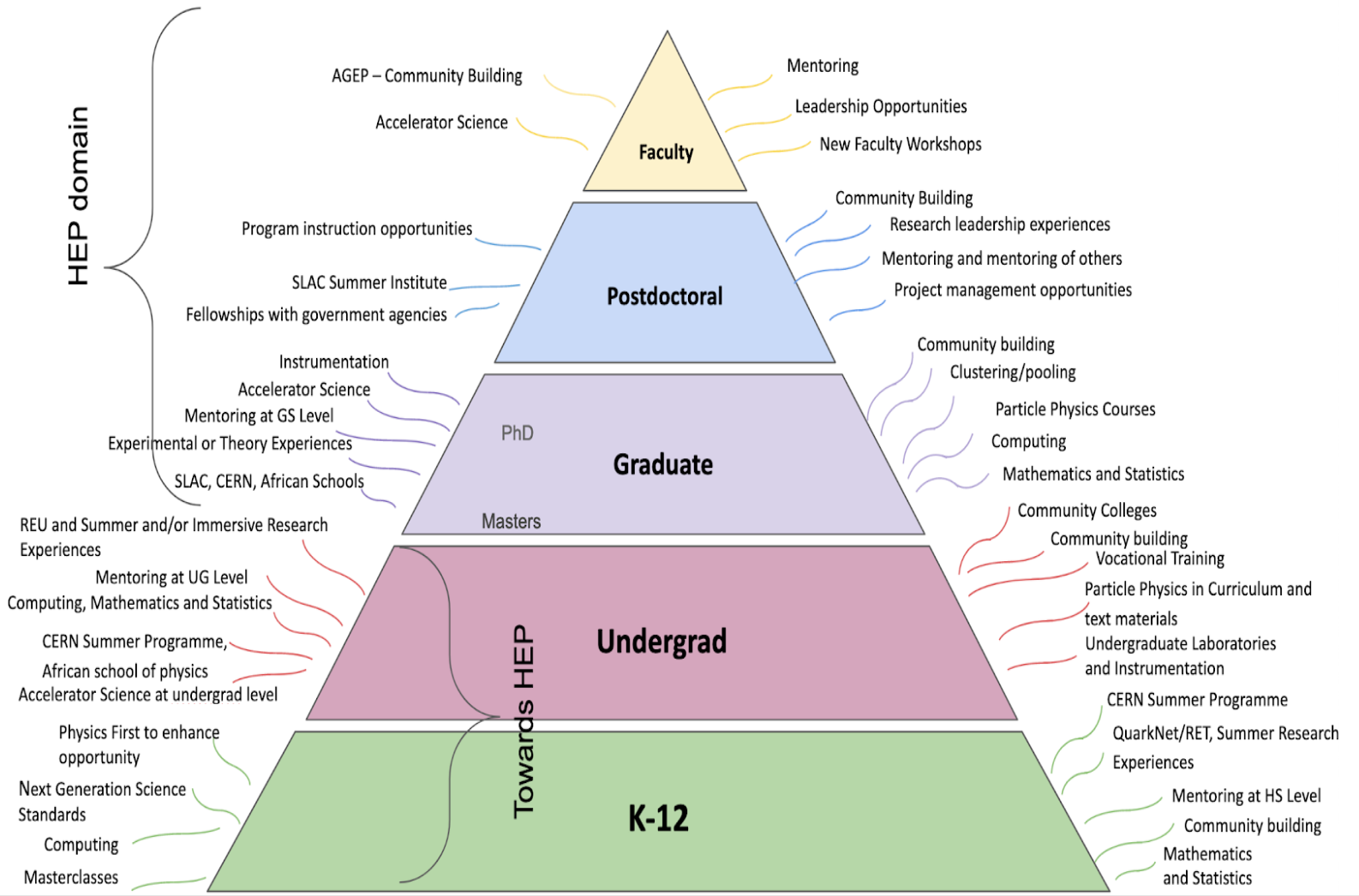}
\hfill
\end{center}
\caption{A Schematic Representation of Particle Physics Education}
\label{fig:Pyramid}
\end{figure}

The PE group framed its studies according to the pyramidal scheme displayed in Figure~\ref{fig:Pyramid}. This diagram rises from the relatively large number of K-12 science students at the base up to the small apex of faculty-level physicists. The work was organized into four groups, each of which produced a contributed paper presenting detailed examinations of physics education challenges and opportunities at each level. The working group contributed papers are:
\begin{itemize}
    \item Opportunities for Particle Physics Engagement in K-12 Schools and Undergraduate Education~\cite{https://doi.org/10.48550/arxiv.2203.10953};
    \item Transforming US Particle Physics Education: A Snowmass 2021 Study~\cite{https://doi.org/10.48550/arXiv.2204.08983};
    \item Broadening the Scope of Education, Career and Open Science in HEP~\cite{https://doi.org/10.48550/arxiv.2203.08809};
    \item The Necessity of International Particle Physics Opportunities for American Education~\cite{https://doi.org/10.48550/arXiv.2203.09336};
\end{itemize}

%%%%%%%%%%%%%%%%%%%%%%%%%%%%%%%%%%%%%%%%%%%%%%%%%%%%%%%%%%%%%%%%%%%%%%%%%
%%%%%%%%%%%%%%%%%%%%%%%%%%%%%%%%%%%%%%%%%%%%%%%%%%%%%%%%%%%%%%%%%%%%%%%%%%%

\subsection{Particle Physics Engagement Opportunities in K-12 Education}
The first working group found that partnerships between academia and K-12 teachers and students are effective in nurturing early student interest in math and science. It is important to provide a broad exposure to the different STEM fields at all levels in order to develop properly their scientific literacy. The key recommendation for forming these partnerships is: 

\begin{itemize}
    \item {\bf At the local level, form collaborative communities (”fora”) of academics of all backgrounds (physicists, engineers, technicians and K-12 teachers). (CEF04 Recommendation 1)}
\end{itemize}

\begin{quote}
    To create these fora, a minimal amount of support for coordination and logistics will need to be available. Then to function well and avoid isolation, they need to be supported by a nationally, or even internationally, organized online repository for sharing resources. To be sustainable, it will be important to have a steady source for continued support, which could come from colleges, universities or institutes, but might also come from or be supplemented by outreach support in the form of research grants. Another important ingredient for sustainability is that the efforts of the fora participants are appropriately and regularly recognized. (CEF04 Recommendations 2,3)
\end{quote}

\subsection{Educational Opportunities at the Undergraduate, Graduate and Postdoctoral Levels}
Education and training specific to particle physics generally begins with college undergraduates, and continues during the graduate student and postdoctoral researcher stages, and usually takes place at universities and laboratories. An online survey of the US HEP community was conducted to gain insight on students’ experiences during this phase. On one hand, respondents indicated that this is the point at which students are expected to gain the skills necessary to launch a career in HEP or related fields. However, they also expressed that the training they received through formal education did not match the skills needed to succeed in physics. In contrast, many of the most important skills they routinely use in their research were learned “on-the-job.” To better prepare students for physics careers, updates to formal university training is recommended~\cite{https://doi.org/10.48550/arxiv.2203.10953}.

\begin{itemize}
    \item {\bf University degree programs should normalize training for particle physics and a broad range of STEM careers through inclusion of appropriate formal courses and career mentoring. (CEF04 Recommendations 4,5)}
\end{itemize}

\begin{quote}
    There is a need for graduate programs in particle physics to provide formal courses with strong grounding in particle physics and mathematics, but also computation, statistics and instrumentation. This course instruction will benefit student careers in physics, industry or education, because the students will not be forced to resort prematurely to self-teaching or peer learning alone. Universities should provide undergraduate students with a more complete picture of what particle physics researchers do. By presenting a realistic view of common career paths post baccalaureate and postgraduate school, students will be better prepared to pursue options including theoretical and experimental positions as well as non-academic careers.
    
    The survey data was limited due to minimal undergraduate participation. This low response was due in part to a lack of connection information for the undergraduate demographic. Support from professional societies and Physics Departments could provide opportunities to strengthen connections and networking for undergraduate students with HEP community activities, and to develop a future survey focused on undergraduate participation. (CEF04 Recommendation 6)
\end{quote}

The physics community and HEP in particular, tends to be fairly “PhD driven.” It is worth investigating whether a Masters degree program in particle physics could fill an important career path need in our field. Masters programs often have the opportunity for more cross-disciplinary work in adjacent fields such as engineering or computer science. Professional level Masters degrees could attract students working in private companies pursuing career advancement through continuing education, and build stronger bi-directional ties between HEP and the industry sector. PhD programs in physics often present strong structural barriers to many students from traditionally underrepresented groups, while a Masters degree could offer an intermediate and more achievable goal, and perhaps lead to greater diversity in HEP. 

\begin{itemize}
    \item {\bf Universities, especially non-research universities, should consider setting up Masters Degree programs in particle physics and related areas, such as hardware and software technology for Big Science experiments. (CEF04 Recommendation 7)}
\end{itemize}

\subsection{Collaborative Opportunities Across Academia}
An important challenge for HEP is the need to broaden and diversify the pool of talent and expertise drawn to the field. A crucial part of the solution to this problem is to build more collaborations between groups at the R1 research institutions that traditionally represent the vast majority of the HEP community, with R2 institutions, Predominantly Undergraduate Institutions (PUI), and Community Colleges (CC).

\begin{itemize}
    \item {\bf Expand the benefits of faculty collaboration and research opportunities across the broad spectrum of academia and give equivalent opportunities for all in technical and scientific leadership on projects, with appropriate recognition for contributions. (CEF04 Recommendation 8)}
\end{itemize}

\begin{quote}
    A study of new models of collaboration or cooperation that would allow R2/PUI/CC faculty and their students to participate effectively in experiment collaborations could help address the challenges of teaching loads, student training and funding availability that directly impact our non-R1 institution colleagues ability to fully contribute to projects. Making data and analysis platforms broadly accessible will benefit student access and participation. The HEP community should embrace the value of Open Science by defining the scope of making our data and resources publicly available, and the hardware, software and person-power costs associated with such implementation. (CEF04 Recommendations 9,10)
\end{quote}

Fields such as instrumentation, computation, and machine learning have become critical components of the HEP enterprise. However, career paths at the intersection of particle physics and these specialized fields are not very clear or easy to navigate, nor universally recognized as “physics” work. Improving this situation would simultaneously address pipeline and retention issues within HEP, and equipping colleagues for careers outside the field.

\begin{itemize}
    \item {\bf Qualification for HEP faculty jobs should not be based solely on physics analysis but rather expanded to include computing, software and/or hardware contributions. (CEF04 Recommendation 11)}
\end{itemize}

\subsection{International Opportunities for Particle Physics Education}
High Energy Physics is conducted through global international collaborations. These diverse partnerships enrich the intellectual environment of our field. As such, training in international collaboration throughout the educational process will facilitate more productive integration of talent and resources in future projects:

\begin{itemize}
    \item {\bf U.S. based pre-university particle physics collaborations should expand collaboration with international partners. (CEF04 Recommendation 12)}
\end{itemize}

\begin{quote}
    Collaborations such as QuarkNet and other outreach programs, such as the International Particle Physics Outreach Group (IPPOG), the CERN Beamline for Schools (BL4S) and Teacher summer school programs in Europe, should partners with counterparts in the developing world, such as the African School of Fundamental Physics and Applications. Participation in the Global Cosmics portal should be enhanced by developing low-cost cosmic ray detectors for educational use.
\end{quote}

\begin{itemize}
    \item {\bf Student exchange programs should be fostered and supported. (CEF04 Recommendation 13)}
\end{itemize}

\begin{quote}
    These programs include the NSF Research Experience for Undergraduates (REU), which funds participation of U.S. students in the CERN Summer Student program, and the DoE-INFN summer student exchange program between the U.S. and Italy. Where possible these should be extended, in particular with student exchange programs and summer schools in developing countries, such as the African School of Fundamental Physics and Applications.
\end{quote}

\section{CEF05: Public Education and Outreach}
\label{sec:CEF05}

\renewcommand{\cefgroup}{5}

The CEF05 working group focused on enabling members of the physics community to effectively communicate about scientific research through public engagement.

Thirteen LOIs were tagged to the Public Education and Outreach topical group. Some of them were consolidated and developed into contributed papers in Section~\ref{sec:CEF03} about diversity, equity and inclusion~\cite{MarginalizedCommunities,StrategiesToEnhanceAcceleratorWorkforce} and Section~\ref{sec:CEF04} about physics education~\cite{https://doi.org/10.48550/arxiv.2203.10953, https://doi.org/10.48550/arxiv.2203.08809,  https://doi.org/10.48550/arXiv.2203.09336}. Other LOIs were condensed into two contributed paper topics, namely:
\begin{enumerate}
    \item The need for structural changes to create impactful public engagement in US particle physics”~\cite{https://doi.org/10.48550/arxiv.2203.08916};
    \item “Particle Physics Outreach at Non-traditional Venues”~\cite{https://doi.org/10.48550/arxiv.2203.09585}.
\end{enumerate}
CEF05 collected input through a variety of methods. In addition to reviewing the LOIs, the group invited experts to their regular meetings for discussion and conducted a survey of the physics community. 

The majority of the survey's 358 respondents said they had participated in outreach activities. They mentioned that they were discouraged from participating in public engagement because they did not have enough time and because it generally did not benefit their careers. They were motivated to participate, though, because they wanted to reach underserved groups, to show openness or explain the scientific method to the public, to share their enthusiasm, and to inspire the next generation of physicists. In their engagement, they used storytelling and shared their own reasons for pursuing physics. Respondents said the best way to get involved in public engagement was to start small and gain experience by finding and plugging into established engagement programs. Details on the group activities are compiled in the topical group report~\cite{https://doi.org/10.48550/arxiv.2210.00983}. 

The group identified several structural and cultural barriers to participation in public engagement. To remove those barriers and encourage physicists to engage the public, the group made specific recommendations aimed at research groups, experimental collaborations, conferences, universities and colleges, national laboratories, OSTP, Congress, DOE, NSF, private foundations, AAAS, APS, and DPF~\cite{https://doi.org/10.48550/arxiv.2203.08916}.

In general, the group recommends:
\begin{itemize}
\item Providing or financially supporting training in effective public engagement
\item Supporting the creation of public engagement programs that scientists can participate in
\item Codifying the importance of public engagement in official documents such as: 
    \begin{itemize}
    \item Laboratory contracts
    \item Faculty handbooks
    \item Professional society strategic plans
    \item Experimental collaboration constitutions
    \item Merit criteria used by institutions that fund research
    \end{itemize}

\item Considering public engagement along with activities such as service and teaching in: 
    \begin{itemize}
    \item Hiring
    \item Tenure
    \item Promotion
    \item Other reviews
    \end{itemize}
\item Funding public engagement work as part of grant proposals 

\item Incorporating public engagement into conferences and meetings in the form of
    \begin{itemize}
    \item Plenary talks
    \item Parallel sessions
    \item Public lectures
    \item Training opportunities for conference participants
    \item Public engagement opportunities for conference participants 
    \end{itemize}
\item Recognizing and rewarding scientists who contribute to public engagement efforts
\end{itemize}

The group also recommends individual scientists encourage others, including peers, mentees and students, by participating in public engagement and discussing its importance.

For the next Snowmass process, CEF05 recommends a shift in focus from "public outreach" to "public engagement": two-way interactions that ensure mutual learning, which goes beyond the acquisition or transmission of knowledge and includes the understanding of perspectives, worldviews and socioeconomic backgrounds. Some innovative ideas on public engagement are discussed in Ref.~\cite{https://doi.org/10.48550/arxiv.2203.09585}, and details are provided in the topical group report~\cite{https://doi.org/10.48550/arxiv.2210.00983}. 
 
The group recommends updating the topical group name from “Public Education and Outreach” to simply “Public Engagement” to reflect this shift in priorities, and also to clear up confusion between the goals of CEF05 and CEF04, the topical group focused on education. 

Public engagement can help recruit and retain scientists from diverse backgrounds, thus improving diversity as discussed in Section~\ref{sec:CEF03}. Therefore, learning how to reach members of marginalized communities via public engagement is essential. Public engagement conducted without proper preparation can be counter-productive and harmful~\cite{dawson2014not}.  Working toward addressing the needs of the intended community must be the objective, achievable through building relations and inclusion~\cite{MarginalizedCommunities}. A detailed checklist of questions for institutions to address when preparing to engage marginalized communities is mentioned in Section~\ref{sec:CEF03} and recalled here.  

Consider the audience:
\begin{itemize}
    \item Who specifically are we hoping to reach with this event? Why are we hoping to reach these communities?
    \item How can we plan this event to make it maximally beneficial to these communities? What elements of this plan can we continue to use in other events?
    \item What are the best ways to communicate about this event with members of these communities? Can we continue going to those same channels to communicate about other events?
    \item Have we created a process by which we take time to evaluate the success of the event after it concludes? 
    \item What metrics (both qualitative and quantitative) will we use? Which of these metrics will we continue to use in evaluating other events?
    \end{itemize}
Identify and remove barriers:
\begin{itemize}
    \item Are there logistical barriers (e.g. time of day, day of the week, public transportation access, affordability, safety concerns, financial barriers) to our events that make them inaccessible to these communities? What will we do to address these barriers?
    \item Have we allowed adequate lead time and budget to make this event accessible to all members of these communities, including those with disabilities? Have we identified partnership or staffing needs required to make the event accessible?
    \end{itemize}
Value partnerships:
    \begin{itemize}
    \item What members of these communities will make good partners in this event? Have we made sure they’re involved in planning the event? Have we secured an adequate budget to support fair compensation for our partners as co-creators of the event, prior to requesting their labor?
    \item Do any members of these communities work for our institution? If they do, do they work in roles with decision-making power (e.g. managerial positions), or do they work primarily in service roles? If members of these communities do not work at our institution, or work only in lower-level positions, is our institution making any effort to change this?
    \item Are members of these communities who work for our institution participating in this event? If so, are they receiving the support they need to take on this effort and fulfill their other job duties? Do they have decision-making power over the planning and execution of the event? Are they being fairly compensated and recognized for their efforts?
    \end{itemize}
Build lasting relationships:
\begin{itemize}
    \item Is this event a part of a larger effort to build relationships with members of these communities? If so, what is the long-term plan? Who will be responsible for enacting it?
    \item Are there ways in which our institution is causing harm to members of these communities? If so, how is our organization working to change this?
    \item How are representatives of our institution involved in these communities outside of this event? Are there ways our institution can work with members of these communities on their priorities, even ones that do not directly benefit our institution?
    \end{itemize}

CEF05 recommends finding ways to implement the structural changes needed for improved public engagement. The group further recommends that the physics community build lasting relationships with marginalized communities through public engagement; this will contribute to improve diversity in HEP as discussed in Section~\ref{sec:CEF03}. Finally, CEF05 recommends that the American Physical Society’s Division of Particles and Fields monitor progress toward these goals leading up to the next Snowmass process.

\section{CEF06: Public Policy and Government Engagement}
\label{sec:CEF06}

\renewcommand{\cefgroup}{6}

The topical group CEF06: Public Policy and Government Engagement (PPGE) was tasked with conducting a review of all current interactions between the HEP community and government offices and individuals. This enterprise includes identification of consensus positions on policies with direct impact on our field, development of unified messages from HEP to those determining and implementing policy, and creation and deployment of tools and resources to effectively communicate those messages in a manner resulting in positive policy outcomes. Those working in CEF06 identified areas of HEP government engagement that are missing or in need of improvement, and developed recommendations to address these opportunities. Three contributed papers produced by CEF06 document this work:
\begin{itemize}
    \item Congressional Advocacy for HEP Funding~\cite{https://doi.org/10.48550/arxiv.2207.00122};
    \item Congressional Advocacy for Areas Beyond HEP Funding~\cite{https://doi.org/10.48550/arxiv.2207.00124};
    \item Non-congressional Government Engagement~\cite{https://doi.org/10.48550/arxiv.2207.00125}.
\end{itemize}
and details of the analysis, synthesis, and recommendations based on those papers is presented in  Ref.~\cite{https://doi.org/10.48550/arxiv.2209.09067}. 

\subsection{HEP Funding and Advocacy Organization}
Over the past few decades, HEP communication with government has largely focused on advocacy for strong federal budget support for HEP, which comes almost exclusively through the Department of Energy (DOE) Office of Science (OS) and the National Science Foundation (NSF). Funding of federal government programs is an extremely complex cyclical process; however there are three basic steps that provide target points for our advocacy. The first is the creation of the annual President’s Budget Request (PBR), which is formulated by the Office of Management and Budget (OMB), with advice from the Office of Science and Technology Policy (OSTP), which works closely with DOE and NSF. The second step is for Congress to pass a budget, which sets topline numbers for funding each major area of government spending. Although the Budget Committees create the budget, it is informed and guided by individual authorization bills, which specify what Congress may spend money on, and these bills come out of authorization committees. Getting specific language supporting HEP programs into authorization bills greatly increases the likelihood of positive funding outcomes. The third major step is appropriations. Appropriations Committees make the decisions on the actual yearly allocation of funds to all government agencies and programs within the constraints of the Congressional budget topline numbers.

The DOE OS and NSF receive HEP program planning advice from a federal advisory committee, the High Energy Physics Advisory Panel (HEPAP). HEPAP has a subpanel known as the Particle Physics Project Prioritization Panel (P5), which produces reports detailing long-range strategic plans for US HEP that are largely based on studies resulting from the Snowmass community planning process. The most recent P5 report from 2014 has served as the core of our field’s message and advocacy to government for nearly a decade. The effectiveness of our messaging and advocacy over this time is indicated by the fact that DOE funding of HEP has grown by 36\%, or roughly \$300M since 2015. This advocacy is carried out jointly by the Fermilab Users Executive Committee (UEC), US-LHC Users Association (USLUA), and SLAC Users Organization (SLUO) with help from the American Physical Society Division of Particles and Fields (APS DPF). However, this group does not have the mandate nor resources to address any aspects of engagement with the government beyond federal funding advocacy (and not enough even to sustain fully that activity). Some new structure must be put in place to broaden HEP’s engagement with government to effectively work for policies to strengthen our field.

\begin{itemize}
    \item {\bf Representatives of APS DPF, HEPAP, and the user groups, as appropriate, should have dedicated discussions to determine what actions can be taken to advance the recommendations outlined in this report~\cite{https://doi.org/10.48550/arxiv.2209.09067}. (CEF06 Recommendation 1)}
\end{itemize}
    
\subsection{Message Unity Around P5}
Advocacy for the 2014 P5 plan has been very successful, leading to a current DOE budget for HEP in excess of \$1B. Critical elements of that success are that P5 represented a single comprehensive plan for the entire US HEP program that had community-wide buy-in, and this led to one unified message that our field delivered to Congress and the Executive Branch. Prior to the 2014 P5, our messaging was fragmented, with people inside and outside the HEP community bringing their own takes on the HEP program to policymakers that were inconsistent with our community-organized advocacy. HEPAP and the 2023 P5 have to lead the effort to build a consensus message around the new P5 plan. This has to include both educating the community and the government about the new plan, and for the updates and changes to the plan that will inevitably occur.

\begin{itemize}
    \item {\bf HEPAP should build community unity around the 2023 P5 plan and develop a clear messaging strategy spanning the next 10 years. (CEF06 Recommendation 2)}
\end{itemize}

\begin{quote}
    Building consensus will require short-term steps related to the drafting and roll-out of the P5 plan. There must be ample opportunity (outside of HEPAP meetings) for internal presentation and community feedback on the draft plan before its release. Once the plan has been finalized, P5 will need to launch an “education campaign” to communicate details about the plan to HEP community members and other stakeholders such as the funding agencies and policy makers. (CEF06 Recommendations 2.1-2.2)

    Long-term actions to maintain message unity consist largely of ensuring good communication. Each year since 2014, the P5 chair has produced a one-page status report that has proved invaluable for Congressional advocacy among other things. A formal commitment should be made to continue producing those reports. More detailed regular reports and feedback opportunities for the community on P5 plan implementation progress, modifications, and impacts will be crucial for keeping the field united behind the plan. (CEF06 Recommendations 2.3-2.6)
    
    Because P5, by definition, is focused exclusively on projects, it has never before directly addressed issues of community engagement, and it is not structurally set up to do so. However, community engagement issues have become extremely important to the healthy functioning of our field, including our projects. P5 will need to explicitly consider relevant community engagement concerns in its work in order to keep the field unified. (CEF06 Recommendation 2.7)
\end{quote}

\subsection{Congressional Advocacy for HEP funding}
The community’s advocacy for federal funding support of HEP largely consists of the annual trip to Washington, DC, during which a group of our colleagues meet with as many Congressional offices as possible, as well as with OMB, OSTP, DOE, and NSF. The ‘DC Trip’ is organized by UEC, USLUA, and SLUO  each spring to fall between the release of the PBR and the markup of appropriations bills.  These users organizations are composed of elected representatives of our community, and the roster of trip attendees is intentionally selected to broadly represent the entire field. However, the users groups are not fully representative of the field, so the formation of a new broader “HEP Congressional advocacy” group should be considered. The DC  Trip has grown dramatically over the past two decades. Around 2004, roughly 25 attendees visited about 150 offices. With increased funding support, by 2019 almost 70 attendees visited all 541 Congress members’ offices as well as 8 subcommittee staff and the Executive Branch offices. 

\begin{itemize}
    \item {\bf Representatives of UEC, SLUO, USLUA, and APS DPF should facilitate discussions to consider the formation of a more formal “HEP Congressional advocacy" group to assume responsibility for organizing the annual advocacy trip to Washington, D.C. (CEF06 Recommendation 3)}
    \item {\bf The HEP Congressional advocacy group should continue to support, and should aim to grow, the annual HEP community-driven advocacy activities. (CEF06 Recommendation 4)}
\end{itemize}

\begin{quote}
    The DC Trip requires a huge organizational effort that has been made possible by the development of a number of tools and resources. Foremost among these is the Washington-HEP Integrated Planning System (WHIPS), which is a framework to automate most of the planning, execution, and documentation logistics. WHIPS compiles information on Congressional districts, offices, and committees, tracks all past and future meetings, and uses data on trip attendees’ personal, work, and family connections to districts to assign attendees to specific Congressional offices. It is the key tool that has enabled HEP advocacy visits to achieve complete coverage of Congress. There is also a twiki repository of trip information and an HEP funding and grant database, both of which could be expanded to include more granular district-level information. These resources have all been developed and maintained by the volunteer effort of a handful of early career colleagues without permanent positions, some who are no longer within the field. A more permanent plan for further maintenance, development and sustainability of these and new tools must be implemented to ensure their continued availability. (CEF06 Recommendations 4.1-4.2)
    
    One question colleagues often ask about HEP advocacy is “How do we know it is effective?” What are the diagnostics and metrics that we use to measure the benefit of the advocacy efforts? WHIPS can track information such as Congress members’ voting records on specific legislation and signatures on Dear Colleague letters, and even match those members’ activities to the HEP trip attendees who visited those offices. However, long-term collection and analysis of the data will require resources beyond current trip planning and participation. (CEF06 Recommendation 4.3) 
    
    There are many professionally-produced communication materials created for the DC Trip. These documents convey our advocacy messages to government offices, and many also serve to share different aspects and benefits of HEP to other audiences. They have been developed by the user organizations in concert with the Fermilab Office of Communication, DOE, and the P5 chair, but there is no guarantee that those groups and individuals will be able to continue providing that support. Investments in maintaining those production partnerships should continue. (CEF06 Recommendation 5)
    
    Training materials have also been created to prepare community members for their participation in the DC Trip, and this training has been crucial for enhancing the professionalism and effectiveness of our advocacy. These materials must be regularly updated and deployed. In addition, making them available to the wider HEP community will enable expansion of our advocacy and help ensure unified messaging. Further inreach efforts to inform the field about our advocacy through more frequent talks and annual reports would also help achieve these goals. (CEF06 Recommendations 4.4-4.5)
    
    Some specific aspects of the DC Trip require specialized knowledge and experience that currently is held by a small number of long-term participants. These include organizing meetings and building relationships with OMB, OSTP, and Congressional subcommittee staff, and the development and use of WHIPS and other tools. Through documentation, this knowledge base needs to be expanded to more participants and archived for future leaders. (CEF06 Recommendations 4.6)
    
    Because the users groups have organized the DC Trip, participants have skewed to experimental Energy and Intensity Frontier colleagues. Although efforts are made to achieve broad representation of the community on the trip roster, more needs to be done to ensure representative participation from segments of our field such as Theory and Computation, from colleagues at all career stages, and from underrepresented groups. (CEF06 Recommendation 4.7)
    
    Finally, opportunities for year-round advocacy should be pursued to engage specific offices (including Congressional local district offices) at other key points in the budget cycle. These opportunities must be weighed against the additional resource costs that would be required. (CEF06 Recommendation 4.8)
\end{quote}

\subsection{Non-Congressional Advocacy}
As was mentioned previously, the DC Trip includes meetings with staff from OSTP and OMB. These are the people who provide policy and budgetary guidance concerning science funding to the formulation of the PBR. These meetings are opportunities for HEP to convey the priorities of our field to the Administration, and also for us to learn about the Administration’s science priorities. The materials for and timing of the DC Trip are chosen primarily for Congressional advocacy. These choices are not necessarily optimal for Executive branch advocacy. For example, our meetings with OSTP and OMB take place immediately after the completed PBR is released. The potential impact of these meetings is very high, and could be maximized with materials, messages, and timing specifically targeted for these offices. 

\begin{itemize}
    \item {\bf The HEP Congressional advocacy group should work to improve HEP community engagement of the executive branch, especially OMB and OSTP. (CEF06 Recommendation 8)}
\end{itemize}

No HEP-wide advocacy efforts exist that are directed to state or local governments. However, there are state and local engagement efforts carried out between individual facilities and the communities in which they are located. These include the Fermilab Community Advisory Board which provides community input and feedback to Fermilab regarding its programs and projects, and similar situations exist with Berkeley Lab and SURF. In all cases, these engagements have been mutually beneficial to the facilities and their communities. 

\begin{itemize}
    \item {\bf The HEP Congressional advocacy group and APS DPF executive committee should facilitate discussions to explore the potential advantages to systematic engagement of local and state governments. (CEF06 Recommendation 9)}
\end{itemize}

\subsection{Advocacy for Issues Beyond HEP Funding}
All of the current HEP community-wide advocacy is directed toward the support and growth of federal funding for HEP. There are many other policy issues not directly tied to HEP funding that nevertheless impact our colleagues and programs of research. Most are or can be addressed with federal legislation. Among these are DEI concerns about limited access to national research facilities for non-R1 institutions, visa and immigration policies that present barriers to foreign scientists wishing to study at or visit US institutions, and balancing research security with open collaboration. While some of these issues are referenced in our DC Trip materials within the context of supporting particle physics, our advocacy infrastructure does not have the resources, procedures, or mandate to advocate for specific non-funding policy positions. Conversations throughout Snowmass on this type of advocacy yielded no consensus between the desire of some to leverage our funding advocacy infrastructure for these issues, and the concerns of others that consensus building on policy issues and potential negative impacts to the field would prove problematic. However, there are external groups with larger resources, constituencies and infrastructure that we have access to for broader advocacy. Among these are the APS, American Institute of Physics (AIP), and American Association for the Advancement of Science (AAAS). All of these groups employ government relations staff, and possess the resources and experience to mobilize advocacy for non-funding issues. They also run very active Congressional Fellowship programs which offer opportunities for HEP community members to participate in much more direct and deeper government engagement on policy. These opportunities should be much more widely promoted to the HEP community.

\begin{itemize}
    \item {\bf The HEP Congressional advocacy group and APS DPF executive committee should identify an existing community group or create a new one to take ownership of strengthening connections between the HEP community and science and physics societies, including APS, AIP, and AAAS. (CEF06 Recommendation 6)}
\end{itemize}

\subsection{Engagement with Funding Agencies}
Another area of government engagement by the HEP community that needs to be improved is direct communication with the funding agencies, DOE and NSF. There are some communication channels that exist, but there are issues with each that prevent them from adequately serving the community. Foremost among the challenges to open communication is the power dynamic between the funding agencies and individual scientists. Groups like HEPAP and the Community of Visitors are explicitly directed to serve as communication channels between the community and agencies, but their memberships are appointed and skewed toward senior colleagues, presenting somewhat of a barrier to younger colleagues feeling comfortable sharing feedback. This is exacerbated by the fact that HEPAP meetings usually have Congressional and/or Executive government officials in attendance. Meetings between DOE/NSF program managers and individual PIs or groups, as well as grant reviews also provide opportunities for communication, but suffer from lack or participation from early career scientists and concerns that negative feedback could have a negative impact on grant applications. None of these channels offers anonymous communication. 

\begin{itemize}
    \item {\bf DOE and NSF should improve existing channels and create new ones, as necessary, to enable HEP community feedback to the funding agencies. (CEF06 Recommendation 7)}
\end{itemize}

\begin{quote}
    There are actions that DOE and NSF could take to remove barriers created by the power dynamic between the agencies and researchers, particularly those in early career stages or from marginalized groups. Chief among these are creating anonymous feedback channels and partnering with community leaders such as the users groups, DPF executive committee and collaboration spokespeople to advertise and encourage the use of communication paths. A particular topic of concern that was frequently expressed is the need to open channels of communication regarding details of the granting process and how that process could be improved. (CEF06 Recommendations 7.1-7.3)
\end{quote}

\section{CEF07: Environmental and Societal Impacts}
\label{sec:CEF07}

\renewcommand{\cefgroup}{7}

This topical group focused on ideas and projects related to how particle physics research impacts society and the environment. Examples of impacts on society include collaboration between particle physics research facilities and indigenous communities related with the land host facilities, or ethical usage of software tools in particle physics research. Examples of impacts on the environment range from the local environment of a research facility (pollution, regional development, international visibility, etc.) all the way to the carbon footprints of particle physics research (experiments, facilities, institutions, etc.). Looking at the long time scale of some particle physics experiment proposals, consideration of the implications of climate change and of the various commitments on carbon emissions reductions by host countries is of paramount importance to ensure the success of particle physics research in the future. Ideas and suggestions on all of those issues were encouraged within this topical group.

Ultimately, five contributed papers were developed and submitted this group, on:
\begin{itemize}
    \item environmental impacts of particle physics~\cite{https://doi.org/10.48550/arxiv.2203.12389, https://doi.org/10.48550/arxiv.2203.07622, https://doi.org/10.48550/arxiv.2203.07423};
    \item interactions of different laboratories with their local communities~\cite{https://doi.org/10.48550/arxiv.2203.07995};
    \item non-nuclear proliferation~\cite{https://doi.org/10.48550/arxiv.2203.00042}.
\end{itemize}

\subsection{Environmental Impacts of Particle Physics}
Particle physics activities include construction and operation of large-scale facilities, detectors, and computing farms, and travels for various types of physics engagements. Doing particle physics impacts on the environment, and this must be considered in the global context of climate change as discussed in Ref.~\cite{https://doi.org/10.48550/arxiv.2203.12389}. Future progress in the field will require the construction of new facilities.  The environmental impacts of facility constructions to advance particle physics research need to be understood in the global efforts to reduce global warming. The projected carbon impacts of just the construction of the main tunnel of the Future Circular Collider (FCC)---or any similar-scale facility---would be comparable to that of a redevelopment of a major city neighborhood; that level of emission will not go unnoticed. The field needs to invest in carbon reduction R\&D and anticipate environmental impact reviews~\cite{https://doi.org/10.48550/arxiv.2203.12389}.  The green ILC is an effort to include carbon reduction in the design, and later in the construction and operation of this machine; a working group has been organized to study efficient design of ILC components and a sustainable ILC City around the laboratory. Should this machine be constructed, its design will be adapted, in consultation with local authorities, to offset excess carbon emission~\cite{https://doi.org/10.48550/arxiv.2203.07622, https://doi.org/10.48550/arxiv.2209.07684}. In Ref.~\cite{https://doi.org/10.48550/arxiv.2203.07423}, the necessity for environmental sustainability in the development of next generation accelerators is articulated; energy-efficient components  and conceptual designs are focus areas for energy efficiency and power consumption in large-scale accelerators. Carbon emissions from greenhouse gases used for detectors and cooling is another area of environmental impact of particle physics activities and mentioned in Ref.~\cite{https://doi.org/10.48550/arxiv.2203.12389}. For future particle physics projects, investment in R\&D is needed for alternative gases---with low global warming potential---for detector operation and facility cooling, without compromising physics performance. Particle physics research relies on large-scale computing; efforts to mitigate computing-related carbon emission should developed, e.g. by optimizing computing-intensive coding and task scheduling~\cite{https://doi.org/10.48550/arxiv.2203.12389}. Greenhouse gas emissions associated to laboratory or university activities are categorized within the scope of direct emissions (from the organization), indirect emissions (electricity, heating, etc.) and other indirect emissions (business travels, commutes, catering, etc.). Across many institutes, much remains to be done to reduce per-capita emissions below the 1 t CO2e per year needed to prevent excessive warming. Travel for physics activities are an essential part of doing particle physics, and aircraft emissions are increasing. It is important to understand what travel are important or necessary and to develop the infrastructures for effective remote engagements. The experiences during the COVID-19 pandemic can serve as  guides to optimize in-person versus remote engagements ~\cite{https://doi.org/10.48550/arxiv.2203.07622, https://doi.org/10.48550/arxiv.2209.07684}. The following guidelines are advanced to reduce the environmental impacts of particle physics activities:
\begin{itemize}
    \item New experiments and facility construction projects should report on their planned emissions and energy usage as part of their environmental assessment which will be part of their evaluation criteria. These reports should be inclusive of all aspects of activities, including construction, detector operations, computing, and researcher activities.
    
    \item US laboratories should be involved in a review across all international laboratories to ascertain whether emissions are reported clearly and in a standardized way. This will also allow other US particle physics research centers (including universities) to use those standards for calculating their emissions across all scopes.
    
    \item Using the reported information as a guide, all participants in particle physics --- laboratories, experiments, universities, and individual researchers --- should take steps to mitigate their impact on climate change by setting concrete reduction goals and defining pathways to reaching them by means of an open and transparent process involving all relevant members of the community. This may include spending a portion of research time on directly tackling challenges related to climate change in the context of particle physics.
    
    \item US laboratories should invest in the development and affordable deployment of next-generation digital meeting spaces in order to minimize the travel emissions of their users. Moreover the particle physics community should actively promote hybrid or virtual research meetings and travel should be more fairly distributed between junior and senior members of the community. For in-person meetings, the meeting location should be chosen carefully such as to minimize the number of long-distance flights and avoid layovers.
    
    \item Long-term projects should consider the evolving social and economic context, such as the expectation of de-carbonized electricity production by 2040, and the possibility of carbon pricing that will have an impact on total project costs.
    
    \item All US particle physics researchers should actively engage in learning about the climate emergency and about the climate impact of particle-physics research. 
    
    \item The US particle physics community should promote and publicize their actions surrounding the climate emergency to the general public and other scientific communities.
    
    \item The US particle physics community and funding agencies should engage with the broader international community to collectively reduce emissions.
\end{itemize}

\subsection{Impact on Local Communities}
Physics engagement with local communities is important to build relations and draw long-lasting community support for particle projects and activities; as noted in Section~\ref{sec:CEF05}, successful and impactful engagements should be foundational rather than transactional. In Ref.~\cite{https://doi.org/10.48550/arxiv.2203.07995}, local community engagement efforts of three laboratories are studied, namely Lawrence Berkeley National Laboratory (Berkeley Lab), Fermi National Accelerator Laboratory (Fermilab), and the Sanford Underground Research Facility (SURF). These laboratories have different local environments, from urban (Berkeley Lab), to suburban (Fermilab), to rural (SURF). In all the three cases, foundational local engagements that promote diversity, communication and lasting relationships, are shown to be mutually beneficial to the community and the laboratory.  We propose the following recommendations for foundational engagements between laboratories and their surrounding communities.

Laboratories should engage with their local communities in order to create awareness about their work and build lasting, positive relationships. Community engagement plays an essential role in local decision-making, building relationships, and important discussions about the implementation of key projects. Large particle physics projects funded by the US Government require an evaluation and mitigation of each project’s potential impacts on the local communities. In addition to satisfying governmental requirements, working alongside their local communities can foster lasting change that broadens the positive societal impacts of particle physics research.
    
Laboratories should have consistent outreach and engagement efforts that provide regular opportunities for feedback to help establish trust. Through its Community Advisory Board, Fermilab offered regularly scheduled meetings to gain feedback from local communities. In addition, SURF ensured its communication with stakeholders at Isna Wica Owayawa was consistent and persistent in order to overcome scheduling and other barriers.
    
Laboratories should promote diversity of membership and collaborative efforts in their outreach initiatives to bring a variety of perspectives to the table and create a better end project. SURF’s work with tribal elders and other leaders in its local community helps ensure perspectives of indigenous populations in the region are represented and reflected in the work of the Sacred Circle Garden. Meanwhile, Fermilab regularly refreshes and expands its CAB membership to ensure it remains representative of the diversity of its suburban area.
    
Laboratories should avoid transactional relationships when developing relationships with stakeholders, and instead focus on approaches that provide value to each entity. Laboratories will be best served by making an extended commitment to working with collaborators over an extended period of time, rather than one-time interactions. Opportunities to receive feedback and consider changes can have lasting impacts on the collaborative efforts. SURF has continued to see improvement in program outcomes with Isna Wica Owayawa using this approach. Berkeley Lab has seen success by utilizing small investments in staff time, small-scale donations, and other resources as a launch pad for lasting collaborations with organizations with shared goals and values.
    
Laboratories should utilize methods that promote honest, two-way communication when engaging in collaborative efforts with stakeholders. All three case studies exemplify the benefits of open communication.The CAB at Fermilab creates a space where local community members and the lab are able to air concerns and discuss solutions. Berkeley Lab ensures that its community engagement interactions provide a space for members of the community and partners to voice their opinions, while Berkeley Lab listens and reflects on the opinions shared. Finally, SURF seeks indigenous perspectives although in some instances, the resulting dialogue can result in uncomfortable conversations. However, by promoting difficult conversations in a safe environment, SURF was able to promote a design for its ethnobotanical garden that was approved by all involved.

\subsection{Impact on Nuclear Non-proliferation}

Detector technologies for neutrino physics can find applications---or benefit from R\&D---in nuclear non-proliferation where detection of reactor anti-neutrinos offer promising reactor monitoring systems that can be remotely operated, robust, non-intrusive and persistent. These ideas are explored in Ref.~\cite{https://doi.org/10.48550/arxiv.2203.00042} and the following recommendations are advanced:

The High Energy Physics community should {\bf continue to engage} in a natural synergy in research activities into next-generation large scale water and scintillator neutrino detectors, now being studied for remote reactor monitoring, discovery and exclusion applications in cooperative nonproliferation contexts.

Examples of ongoing synergistic work at US national laboratories and universities should continue and be expanded upon. These include prototype gadolinium-doped water and water-based and opaque scintillator test-beds and demonstrators, extensive testing and industry partnerships related to large area fast position-sensitive photomultiplier tubes, and the development of concepts for a possible underground kiloton-scale water-based detector for reactor monitoring and technology demonstrations.

Opportunities for engagement between the particle physics and
nonproliferation communities should be encouraged. Examples include the bi-annual Applied Antineutrino Physics conferences, collaboration with US national laboratories engaging in this research, and occasional NNSA funding opportunities supporting a blend of nonproliferation and basic science R\&D, directed at the US academic community.

\subsection{Links with Other Topical groups}
This topical group is interlinked with other Community Engagement Frontier topical groups. For example, the impacts on society include issues involving inclusion and diversity and the impacts on the environment and the sustainability of the field involve components associated with the applications and industry topical group.

\section{Conclusions}
\label{sec:conc}
During Snowmass 2021, participants in the Community Engagement Frontier have attempted to address the importance of engaging members of various communities to generate support for and sustainability of high energy physics. The CEF efforts were organized into seven topical groups and connections with the other Snowmass frontiers were established through frontier liaisons to exchange feedback. Each CEF topical group studied specific focus areas of engagement and their importance to society and the future of our field. The focus areas were Applications and Industry, Career Pipeline and Development, Diversity, Equity and Inclusion, Physics Education, Public Education and Outreach, Public Policy and Government Engagement, and Environmental and Societal Impacts. Topical groups collected and studied inputs from letters of interest, surveys, town hall meetings, workshops, invited expert discussions, and regular working group meetings. These efforts produced thirty-five contributed papers and seven topical group reports containing recommendations to improve engagement between HEP and related communities. A few ideas that were not developed into contributed papers because of lack of person-power are nevertheless noted in the texts. To facilitate maximum impact and efficiency of implementation, the suggested recommendations for action have been directed to different entities within the HEP community, namely government and funding agencies, academic and research institutions, research collaborations, professional societies and individual physicists, and have been organized into overall goals categorized by five target communities for engagement. 

We, the topical group and frontier conveners of CEF, lament the persistently low participation in community engagement. Regrettably, regardless of efforts throughout Snowmass 2021 to motivate participation on cross-cutting CEF issues, the work in this frontier was carried out by a relatively small number of colleagues, most of whom are physicists who also had interests in other physics frontiers that they were largely unable to pursue. Various reasons, some quite understandable, have been advanced to explain this lack of interest. However, until due importance is given to community engagement efforts and mechanisms are implemented for support, encouragement and rewards, no meaningful progress will be achieved in spite of expressed well-meaning intentions. We call upon each specified entities’ members that are serious about improving HEP community engagement to take ownership of the CEF suggestions and act on their implementation within structures developed to foster and gauge progress. We hope that at the next Snowmass, we do not find a repeat of this past decade’s inaction on these issues, but rather that we inherit a vibrant program of HEP community engagement on which to build.

\section*{Acknowledgements}
We appreciate the assistance of Daria Wang (ORISE) in the planning and management of CEF efforts. We thank Alan Stone (DOE Office of Science) for support. We express our appreciation for all of the people that contributed to CEF efforts, e.g. the DPF chair line for support, the authors of contributed papers, the topical experts that shared their thoughts and experiences, the liaisons that helped maintain channels of communications with other frontiers, and the topical group and frontier conveners who carried the heavy loads so diligently and efficiently. 

%%%%%%%%%%%%%%%%%%%%%%%%%%%%%%%%%%%%%%%%%%

%  If you would like to use BibTEX for the bibliography, please feel free to do so.  It is not required.

%  To use BibTeX,

%    1.  uncomment the following two lines,
%    2.  comment out everything below from  \begin{thebibliography}{99}   to \end{thebibliography).
%    3.  create the file  myreferences.bib in this directory, and process this file in the usual way

\bibliographystyle{JHEP}

\bibliography{Engagement/myreferences}

\providecommand{\href}[2]{#2}\begingroup\raggedright\begin{thebibliography}{10}

\bibitem{https://doi.org/10.48550/arxiv.1401.6119}
M.~Bardeen, D.~Cronin-Hennessy, R.~M.~Barnett, P.~Bhat, K.~Cecire, K.~Cranmer
  et~al., \emph{Planning the future of U.S. particle physics (Snowmass 2013):
  Chapter 10: Communication, education, and outreach},  2014.
arXiv:1401.6119.

\bibitem{https://doi.org/10.48550/arxiv.2210.01248}
F.~Fahim, A.~Murokh and K.~Yoshimura, \emph{Summary report of the topical group on
  application and industry},  2022.
arXiv:2210.01248.

\bibitem{https://doi.org/10.48550/arxiv.2209.10114}
J.~Hogan, A.~Karadzhinova-Ferrer and S.~Malik, \emph{Summary report of the
  topical group on career pipeline and development},  2022.
arXiv:2209.10114.

\bibitem{https://doi.org/10.48550/arxiv.2209.12377}
C.~Bonifazi, J.~S.~Bonilla, M.~C.~Chen, Y.~H.~Lin, K.~A.~Assamagan, E.~V.~Hansen
  et~al., \emph{Diversity, equity, and inclusion in particle physics},  2022.
arXiv:2209.12377.

\bibitem{https://doi.org/10.48550/arxiv.2209.08225}
S.~J.~de~Jong, S.~Mahlik and R.~Ruchti, \emph{Summary report of the topical
  group on physics education},  2022.
arXiv:2209.08225.

\bibitem{https://doi.org/10.48550/arxiv.2210.00983}
S.~Demers, K.~Jepsen, D.~Lincoln and A.~Muronga, \emph{Public education and
  outreach},  2022.
arXiv:2210.00983.

\bibitem{https://doi.org/10.48550/arxiv.2209.09067}
R.~Fine and L.~Suter, \emph{Public policy \& government engagement},  2022.
arXiv:2209.09067.

\bibitem{https://doi.org/10.48550/arxiv.2209.07684}
K.~Bloom, V.~Boisvert and M.~Headley, \emph{Report of the topical group on
  environmental and societal impacts of particle physics},
  2022.
arXiv:2209.07684.

\bibitem{https://doi.org/10.48550/arxiv.2203.07328}
G.~Agarwal, J.~L.~Barrow, M.~F.~Carneiro, E.~Conley, M.~E.~de~S.~Pereira, S.~Hedges
  et~al., \emph{Snowmass 2021 community survey report},  2022.
arXiv:2203.07328.

\bibitem{https://doi.org/10.48550/arxiv.2203.15128}
A.~Arai, F.~Fahim, R.~Furubayashi, M.~Garrett, S.~Li, K.~McDonald et~al.,
  \emph{Programs enabling deep technology transfer from national labs},  2022.
arXiv:2203.15128.

\bibitem{https://doi.org/10.48550/arxiv.2203.14706}
J.~Hoff and S.~Memik, \emph{Application-driven engagement with universities,
  synergies with other funding agencies},  2022.
arXiv:2203.14706.

\bibitem{https://doi.org/10.48550/arxiv.2203.08973}
G.~Carini, M.~Demarteau, P.~Denes, A.~Dragone, F.~Fahim, C.~Grace et~al.,
  \emph{Big industry engagement to benefit hep: Microelectronics support from
  large cad companies},  2022.
arXiv:2203.08973.

\bibitem{https://doi.org/10.48550/arxiv.2203.11047}
S.~Boucher, E.~Esarey, C.~Geddes, C.~Johnstone, S.~Kutsaev, B.~W.~Loo et~al.,
  \emph{Transformative technology for flash radiation therapy: A snowmass 2021
  white paper},  2022.
arXiv:2203.11047.

\bibitem{https://doi.org/10.48550/arxiv.2203.10377}
A.~M.~M.~Todd, R.~Agustsson, D.~L.~Bruhwiler, J.~Chunguang, S.~C.~Gottschalk,
  A.~Kanareykin et~al., \emph{Nurturing the industrial accelerator technology
  base in the US},  2022.
arXiv:2203.10377.

\bibitem{https://doi.org/10.48550/arxiv.2203.11665}
S.~Malik, A.~Karadzhinova-Ferrer, J.~Hogan, R.~Bray, R.~Kamalieddin, K.~Flood
  et~al., \emph{Facilitating non-HEP career transition},  2022.
arXiv:2203.11665.

\bibitem{https://doi.org/10.48550/arxiv.2203.11662}
M.~Bellis, B.~Bhattacharya, D.~DeMuth, J.~Hogan, K.~Laureto, S.~Malik et~al.,
  \emph{Enhancing HEP research in predominantly undergraduate institutions and
  community colleges},  2022.
arXiv:2203.11662.

\bibitem{AccessibilityInHEP}
K.~A.~Assamagan, C.~Bonifazi, J.~S.~Bonilla, P.~A.~Breur, M.~C.~Chen, A.~Roepe-Gier
  et~al., \emph{Accessibility in high energy physics: Lessons from the
    Snowmass
  process},  2022.
arXiv:2203.08748.

\bibitem{LifestyleAndPersonalWellness}
T.~R.~Lewis, S.~M.~Simon, C.~Bonifazi, S.~Thais, J.~S.~B.~Castro and
  K.~A.~Assamagan, \emph{Lifestyle and personal wellness in particle physics
  research activities},  2022.
arXiv:2203.08631.

\bibitem{ClimateOfTheField}
E.~V.~Hansen, E.~Smith, D.~Bard, M.~Bellis, J.~Esquivel, T.R.~Lewis et~al.,
  \emph{Climate of the field: Snowmass 2021},  2022.
arXiv:2204.03713.

\bibitem{HEPInAfricaAndLatinAmerica}
K.A.~Assamagan, C.~Bonifazi, J.~S.~B.~Castro, C.~David, C.~Dib, L.~D.~S.~Matias
  et~al., \emph{Why should the U.S. care about high energy physics in Africa
  and Latin  America?},  2022.
arXiv:2203.10060.

\bibitem{MarginalizedCommunities}
K.~A.~Assamagan, O.~M.~Bitter, M.-C.~Chen, A.~Choi, J.~Esquivel, K.~Jepsen
  et~al., \emph{Building a culture of equitable access and success for
  marginalized members in today's particle physics community},  2022.
arXiv:2206.01849.

\bibitem{HowToReadSelectedSnowmassPapers}
A.~K.~Hodari, S.~B.~Krammes, C.~Prescod-Weinstein, B.~D.~Nord, J.~N.~Esquivel and
  K.~A.~Assamagan, \emph{How to read the Snowmass white papers on power dynamics
  in physics, informal socialization in physics training, and policing and
  gatekeeping in STEM},  2022.
arXiv:2203.11523.

\bibitem{PowerDynamicsInPhysics}
A.~K.~Hodari, S.~B.~Krammes, C.~Prescod-Weinstein, B.~D.~Nord, J.~N.~Esquivel and
  K.~A.~Assamagan, \emph{Power dynamics in physics},  2022.
arXiv:2203.11513.

\bibitem{PolicingAndGatekeepingInSTEM}
A.~K.~Hodari, S.~B.~Krammes, C.~Prescod-Weinstein, B.~D.~Nord, J.~N.~Esquivel and
  K.A.~Assamagan, \emph{Policing and gatekeeping in STEM},  2022.
arXiv:2203.11508.

\bibitem{InformalSocializationInPhysicsTraining}
A.~K.~Hodari, S.~B.~Krammes, C.~Prescod-Weinstein, B.~D.~Nord, J.~N.~Esquivel and
  K.~A.~Assamagan, \emph{Informal socialization in physics training},  2022.
arXiv:2203.11518.

\bibitem{ExcellenceAndEquityInPhysics}
E.~Barzi, S.~J.~Gates and R.~Springer, \emph{In search of excellence and equity
  in physics},  2022.
arXiv:2203.10393.

\bibitem{StrategiesToEnhanceAcceleratorWorkforce}
M.~Bai, W.~A.~Barletta, D.~L.~Bruhwiler, S.~Chattopadhyay, Y.~Hao, S.~Holder
  et~al., \emph{Strategies in education, outreach, and inclusion to enhance the
  us workforce in accelerator science and engineering},  2022.
arXiv:2203.08919.

\bibitem{TitleVI}
C.~R.~Service, \emph{Foreign language and international studies: Federal aid
  under Title VI of the higher education act},  2008.

\bibitem{AAASDiversityAndTheLaw}
{American Association for the Advancement of Science (AAAS)}, ``Diversity and
  the {{Law}}.'' https://www.aaas.org/programs/diversity-and-law.

\bibitem{AIPTeamUpWebsite}
{American Institute of Physics}, ``{{TEAM-UP Project}}.''
  \url{https://www.aip.org/diversity-initiatives/team-up-task-force}.

\bibitem{TEAMUPreport2020}
{The AIP National Task Force to Elevate African American Representation in
  Undergraduate Physics \& Astronomy (TEAM-UP)}, \emph{The {{Time}} is {{Now}}:
  {{Systemic Changes}} to {{Increase African Americans}} with {{Bachelor}}'s
  {{Degrees}} in {{Physics}} and {{Astronomy}}},  Tech. Rep. {American
  Institute of Physics} (Jan., 2020).

\bibitem{AAASSeaChange}
{American Association for the Advancement of Science (AAAS)}, ``{{SEA
  Change}}.'' https://seachange.aaas.org.

\bibitem{nsfADVANCEOrganizationalChange}
{National Science Foundation (NSF)}, ``{{ADVANCE}}: {{Organizational Change}}
  for {{Gender Equity}} in {{STEM Academic Professions}} ({{ADVANCE}}).''
  \url{https://beta.nsf.gov/funding/opportunities/advance-organizational-change-gender-equity-stem-academic-professions-advance}.

\bibitem{https://doi.org/10.48550/arxiv.2203.10953}
M.~G.~Bardeen, O.~M.~Bitter, M.~Glover, S.~J.~de~Jong, T.~R.~Lewis, M.~Fetsko
  et~al., \emph{Particle physics outreach to K-12 schools and opportunities in
  undergraduate education},  2022.
arXiv:2203.10953.

\bibitem{https://doi.org/10.48550/arXiv.2204.08983}
O.~Bitter, E.~V.~Hansen, S.~Kravitz, V.~Velan and Y.~You, \emph{Transforming
  U.S. particle physics education: A Snowmass 2021 study},  2022.
arXiv:2204.08983.

\bibitem{https://doi.org/10.48550/arxiv.2203.08809}
S.~Malik, D.~DeMuth, S.~de~Jong, R.~Ruchti, S.~Thais, G.~Fidalgo et~al.,
  \emph{Broadening the scope of education, career and open science in hep},
  2022.
arXiv:2203.08809.

\bibitem{https://doi.org/10.48550/arXiv.2203.09336}
E.~Arce-Larreta, K.~Assamagan, E.~Barzi, U.~Bilow, K.~Cecire, S.~de~Jong
  et~al., \emph{The necessity of international particle physics opportunities
  for American education},  2022.
arXiv:2203.09336.

\bibitem{https://doi.org/10.48550/arxiv.2203.08916}
K.~A.~Assamagan, M.~Carneiro, S.~Demers, K.~Jepsen, D.~Lincoln and A.~Muronga,
  \emph{The need for structural changes to create impactful public engagement
  in US particle physics},  2022.
arXiv:2203.08916.

\bibitem{https://doi.org/10.48550/arxiv.2203.09585}
J.~Cochran, J.~Huth, R.~Jones, P.~Laycock, C.~Lee, L.~Lee et~al.,
  \emph{Particle physics outreach at non-traditional venues},  2022.
arXiv:2203.09585.

\bibitem{dawson2014not}
E.~Dawson, \emph{“not designed for us”: How science museums and science
  centers socially exclude low-income, minority ethnic groups}, {\emph{Science
  education} {\bfseries 98} (2014) 981}.

\bibitem{https://doi.org/10.48550/arxiv.2207.00122}
M.~Carneiro, R.~Diurba, R.~Fine, M.~Gill, K.~Kaadze, H.~Newman et~al.,
  \emph{Snowmass '21 community engagement frontier 6: Public policy and
  government engagement: Congressional advocacy for HEP funding (the "DC
  trip'')},  2022.
arXiv:2207.00122.

\bibitem{https://doi.org/10.48550/arxiv.2207.00124}
R.~Diurba, R.~Fine, M.~Gill, H.~Newman, K.~Pedro, A.~Perloff et~al.,
  \emph{Snowmass '21 community engagement frontier 6: Public policy and
  government engagement: Congressional advocacy for areas beyond HEP funding},
  2022.
arXiv:2207.00124.

\bibitem{https://doi.org/10.48550/arxiv.2207.00125}
R.~Diurba, R.~Fine, M.~Gill, H.~Newman, K.~Pedro, A.~Perloff et~al.,
  \emph{Snowmass '21 community engagement frontier 6: Public policy and
  government engagement: Non-congressional government engagement},  2022.
arXiv:2207.00125.

\bibitem{https://doi.org/10.48550/arxiv.2203.12389}
K.~Bloom, V.~Boisvert, D.~Britzger, M.~Buuck, A.~Eichhorn, M.~Headley et~al.,
  \emph{Climate impacts of particle physics},  2022.
arXiv:2203.12389.

\bibitem{https://doi.org/10.48550/arxiv.2203.07622}
A.~Aryshev et~al., \emph{The International Linear Collider: Report to
  snowmass 2021},  2022.
arXiv:2203.07622.

\bibitem{https://doi.org/10.48550/arxiv.2203.07423}
T.~Roser, \emph{Sustainability considerations for accelerator and collider
  facilities},  2022.
arXiv:2203.07423.

\bibitem{https://doi.org/10.48550/arxiv.2203.07995}
R.~Zens, M.~Headley, D.~Wolf, A.~Markovitz, F.~Dukes, J.~Tang et~al.,
  \emph{Societal impacts of particle physics projects},  2022.
arXiv:2203.07995.

\bibitem{https://doi.org/10.48550/arxiv.2203.00042}
T.~Akindele et al., \emph{A call to arms control: Synergies between
  nonproliferation applications of neutrino detectors and large-scale
  fundamental neutrino physics experiments},  2022.
arXiv:2203.00042.

\end{thebibliography}\endgroup

\end{document}